\documentclass[structabstract]{aa}
\usepackage{graphicx}
\usepackage{txfonts}
\usepackage{natbib}
\usepackage{multirow}
\usepackage{longtable}
\usepackage{pdflscape}
\usepackage{fixltx2e}
\usepackage[colorinlistoftodos,color=blue!20]{todonotes}

\def\4he{$^4$He}

\def\kms{\mathrm{km\,s}^{-1}}
\def\e#1{\times 10^{#1}}

\def\msol{\mathrm{M}_\odot}

\def\up#1{$^{#1}$}

\def\h2{$\mathrm{H}_2$}
\def\so2{$\mathrm{SO}_2$}
\def\mic{\mathrm{ }\mu\mathrm{m}}
\def\spy{\;\msol~\mathrm{ yr}^{-1}}

\begin{document}
	\title{Sulphur-bearing molecules in AGB stars}
   \subtitle{I: The occurrence of hydrogen sulfide}

   \author{T. Danilovich
          \inst{1}\fnmsep\thanks{Postdoctoral Fellow of the Fund for Scientific Research (FWO), Flanders, Belgium}
          \and
          M. Van de Sande\inst{1}
          \and
          E. De Beck\inst{2}
          \and
          L. Decin\inst{1}
          \and
          H. Olofsson\inst{2}
          \and
          S. Ramstedt\inst{3}
          \and
          T. J. Millar\inst{4}
          }

   \institute{Department of Physics and Astronomy, Institute of Astronomy, KU Leuven, Celestijnenlaan 200D,  3001 Leuven, Belgium 
   \and
      Onsala Space Observatory, Department of Earth and Space Sciences, Chalmers University of Technology, 43992 Onsala, Sweden   
      \and
      Department of Physics and Astronomy, Uppsala University, Box 516, 75120, Uppsala, Sweden
      \and 
      Astrophysics Research Centre, School of Mathematics and Physics,
Queen's University Belfast, University Road,
Belfast BT7~1NN, UK
      \\
             \email{taissa.danilovich@kuleuven.be}
             }

   \date{Received  19 May 2017 / Accepted }

 
  \abstract
   {Sulphur is a relatively abundant element in the local galaxy which is known to form a variety of molecules in the circumstellar envelopes of AGB stars. The abundances of these molecules vary based on the chemical types and mass-loss rates of AGB stars.}
   {Through a survey of (sub-)millimetre emission lines of various sulphur-bearing  molecules, we aim to determine which molecules are the primary carriers of sulphur in different types of AGB stars. In this paper, the first in a series, we investigate the occurrence of \h2S in AGB circumstellar envelopes and determine its abundance, where possible.}
   {We have surveyed 20 AGB stars with a range of mass-loss rates and of different chemical types using the APEX telescope to search for rotational transition lines of five key sulphur-bearing molecules: CS, SiS, SO, \so2 and \h2S. Here we present our results for \h2S, including detections, non-detections and detailed radiative transfer modelling of the detected lines. We compare results based on different  descriptions of the molecular excitation of \h2S and different abundance distributions, including Gaussian abundances, where possible, and two different abundance distributions derived from chemical modelling results.}
   {We detected \h2S towards five AGB stars, all of which have high mass-loss rates of $\dot{M}\geq 5\e{-6}\spy$ and are oxygen-rich. \h2S was not detected towards the carbon or S-type stars that fall in a similar mass-loss range. For the stars in our sample with detections, we find peak o-\h2S abundances relative to \h2 between $4\e{-7}$ and $2.5\e{-5}$.}
  {Overall, we conclude that \h2S can play a significant role in oxygen-rich AGB stars with higher mass-loss rates, but is unlikely to play a key role in stars of other chemical types or the lower mass-loss rate oxygen-rich stars. For two sources, V1300~Aql and GX~Mon, \h2S is most likely the dominant sulphur-bearing molecule in the circumstellar envelope.}
   

   \maketitle
%
\section{Introduction}

One of the {latter} stages in the evolution of a low- to intermediate-mass star is the asymptotic giant branch (AGB). The AGB phase is characterised by intense mass loss through a steady stellar wind. This outflow forms a circumstellar envelope (CSE), rich in chemical diversity, in which molecules and dust form. The nature of the CSE is of particular interest as it contains matter that will eventually be returned to the interstellar medium (ISM) and will contribute to the chemical enrichment and evolution of the galaxy. Although most molecules formed in the CSE will eventually be photodissociated by photons from the interstellar UV field, some may be condensed onto dust grains and hence not introduced into the ISM in their atomic form. 

Sulphur is the tenth most abundant element in the universe and one of the essential elements for life on Earth. In the molecularly rich envelopes of AGB stars it is known to form molecules such as CS, SiS, SO, \so2 and \h2S, which vary in abundance depending on chemical type and mass-loss rate. For example, CS is most abundant in carbon stars (compare the abundances found by \cite{Olofsson1993a} for carbon stars and those found by \cite{Lindqvist1988} for oxygen-rich stars), and SO and \so2 are most abundant in oxygen-rich stars \citep{Danilovich2016}. 
Furthermore, sulphur is not formed in AGB stars nor their main sequence progenitors, so the total amount of sulphur-bearing molecules can be constrained based on upper limits from galactic and solar abundances for sulphur. Indeed, \cite{Danilovich2016} began to do this based on a thorough study of SO and \so2 in a small sample of M-type AGB stars.

{It has been suggested that an infrared spectral feature at $30~\mic$ may be due to MgS dust \citep[see, for example,][]{Goebel1985,Begemann1994} and may account for a significant proportion of the available sulphur in carbon-rich AGB stars. However, a more recent study by \cite{Zhang2009} found that, to reproduce the $30~\mic$ feature, a larger quantity of sulphur, in the form of MgS, was required than available in circumstellar envelopes. Hence the feature is likely not due to MgS. This is further emphasised by studies of post-AGB stars, which find significant depletion of refractory elements but minimal depletion of sulphur, indicating that sulphur is not condensed onto dust in large quantities \citep[see, for example,][]{Waelkens1991,Reyniers2007}.}

Chemical models that include sulphur chemistry often take SiS and \h2S to be the parent species, i.e. the molecules in which sulphur is initially contained at the model inner radius \citep{Willacy1997, Cherchneff2006, Cherchneff2012, Agundez2010, Li2016}, especially for oxygen-rich stars. However, the specific choice of parent species and their initial abundances can lead to different outcomes in chemical modelling \citep{Li2016}, including differences in the shape and peak values of various abundance distributions in the CSE. Varying approaches to the chemical modelling also give different results. For example, \cite{Willacy1997} consider the photo-dominated low-density outer part of the CSE, at radii $>10^{15}$~cm, for oxygen-rich stars and, having assumed \h2S to be the key S-bearing parent species, predict a consistent decline in \h2S abundance as it reacts to form other S-bearing species such as CS, SO and \so2. \cite{Li2016} consider a similar region of an oxygen-rich CSE, but assume a much lower inner \h2S abundance --- in their model the majority of the sulphur is initially found in SiS --- and find a slight increase in \h2S before it decreases at a similar rate to the \cite{Willacy1997} models. The models of \cite{Agundez2010} consider a clumpy CSE medium for a range of mass-loss rates, through which UV photons are potentially able to penetrate to the inner regions. Their models of carbon stars, which include inner regions close to the dust condensation region as well as outer regions comparable to the \cite{Willacy1997} and \cite{Li2016} models, find rapid formation of \h2S followed by two stages of destruction, first rapid then more moderate, for carbon-rich CSEs (with their \h2S results for oxygen-rich CSEs not shown).
 \cite{Gobrecht2016} study the innermost regions of an oxygen-rich CSE, closest to the AGB star itself and within the dust condensation region, examining the effects of shocks on molecular abundances and dust condensation. They find a high abundance of \h2S close to the star which drops off rapidly as \h2S is destroyed by various chemical processes. 


\h2S is not commonly detected in AGB stars aside from high mass-loss rate OH/IR stars ($\dot{M} \sim 10^{-4}\spy$). For example, the \textsl{Herschel}/HIFI Guaranteed Time Key Project HIFISTARS observed the \h2S $(3_{1,2}\to2_{2,1})$ line at 1196.012 GHz in 9 M-type AGB stars but only detected it in AFGL~5379, which has been classified as an OH/IR star \citep{Justtanont2012}. \cite{Justtanont2015} detected several \h2S lines in all 8 of the OH/IR stars they observed with \textsl{Herschel}/SPIRE and \textsl{Herschel}/PACS. \cite{Ukita1983} searched for the \h2S $(1_{1,0}\to1_{0,1})$ line at 168.763 GHz in 25 AGB sources and detected it only in OH 231.8~+4.2 (aka the Rotten Egg Nebula), an OH/IR star which may be transitioning to the post-AGB phase. \cite{Omont1993} surveyed a diverse sample of evolved stars for several sulphur-bearing molecules and detected \h2S in 15 sources, including high mass-loss rate AGB stars and OH/IR stars. The sources in which they did not detect \h2S include carbon stars, S-type stars and lower mass-loss rate M-type AGB stars, although they also confirmed the detection of \h2S in CW~Leo, which is the only carbon star for which \h2S has been detected to date, with detections from several studies of the 168 GHz ortho-\h2S line \citep{Cernicharo1987,Omont1993,Cernicharo2000}. {Note, however, that the para-\h2S ($2_{2,0}\to2_{1,1}$) line at 216.710 GHz has not been detected in CW~Leo \citep[see for example the][1~mm survey]{Tenenbaum2010a}, nor has the ortho-\h2S ($3_{3,0}\to3_{2,1}$) line at 300.506 GHz \citep[see for example the][survey]{Patel2011}.}

To better constrain the abundances and distributions of {the most abundant} sulphur-bearing molecules in AGB stars of different chemical types and a range of mass-loss rates, we performed a survey of \h2S, SiS, CS, SO, and \so2 with the Atacama Pathfinder Experiment (APEX\footnote{This publication is based on data acquired with the Atacama Pathfinder Experiment (APEX). APEX is a collaboration between the Max-Planck-Institut f\"ur Radioastronomie, the European Southern Observatory, and the Onsala Space Observatory.}), a 12 m radio telescope located at Llano Chajnantor in northern Chile.
In this series of papers we intend to investigate the occurrence and abundance distributions of these key sulphur-bearing molecules. {Once the abundances are known we intend to conduct a detailed chemical analysis of S-bearing species, including detailed chemical modelling.} In this first study, we present results of the \h2S observations from this survey.  


\section{Sample and observations}\label{obs}

We observed several lines of CS, SiS, SO, \so2 and \h2S in a diverse sample of 20 AGB stars, including 7 M-type stars, 5 S-type stars and 8 carbon stars and covering mass-loss rates from $\sim9\e{-8}\spy$ to $\sim2\e{-5}\spy$. The sample was chosen to cover a range of mass-loss rates across all three chemical types, and was largely drawn from stars covered by the SUCCESS programme \citep{Danilovich2015a}, for which high quality CO observations and mass-loss rates {derived from CO modelling} were already available. Of this sample all but two sources were observed for at least one of three possible \h2S transitions and \h2S was detected in three of these. To expand our sample, an additional three high mass-loss rate M-type sources were added and observed only in the frequency setting containing the \h2S line at 168.763 GHz. \h2S was detected in two of these.

Observations were carried out using the Swedish-ESO PI receiver for APEX \citep[SEPIA Band 5,][]{Billade2012} and the Swedish Heterodyne Facility Instrument \citep[SHeFI,][]{Risacher2006,Vassilev2008}. The data were reduced using the GILDAS/CLASS\footnote{\url{http://www.iram.fr/IRAMFR/GILDAS/}} software package. Following the initial assessment by \cite{Immer2016} of the performance of SEPIA, we assumed a main beam efficiency of $\eta_\mathrm{mb} = 0.68$ for our SEPIA tunings. The well-established main beam efficiencies of  $\eta_\mathrm{mb} = 0.75$ and $\eta_\mathrm{mb} = 0.73$ \citep{Gusten2006} were used for the 216 GHz and 300 GHz observations, respectively. {The beam sizes corresponding to each transition are listed in Table \ref{trans} and the emission is spatially unresolved for all lines observed using APEX}.

The full sample of stars included in the APEX \h2S survey is listed in Table \ref{fullsample}. The lines included in the survey are listed in Table \ref{trans}, along with other available lines from other telescopes. Table \ref{APEXobs} includes all the detected \h2S lines and their integrated main beam intensities and Table \ref{nondet} lists all the observed sources with the RMS noise limit at $1~\kms$ for each observed line. 
We list all non-detections for completion. 

{We encountered no issues with line ambiguity for the 168 GHz and 300 GHz lines; there were no plausible alternative identifications for these lines. The 216 GHz line, which was only tentatively detected in IK~Tau could potentially be an overlap with \up{13}CN. However, the \up{13}CN features around 216 GHz are weak and not even detected in the two carbon stars for which other \up{13}CN groups are seen, around 217 GHz and 218 GHz. Hence, we conclude that the weak emission seen at 216.710 GHz towards IK~Tau is most likely due to \h2S.}

\begin{table*}[tp]
\caption{Basic parameters of surveyed stars.}\label{fullsample}
\begin{center}
\begin{tabular}{lcccccccc}
\hline\hline
Star	&	RA	&	Dec	&	$\upsilon_\mathrm{LSR}$	&	Distance	&		$\dot{M}$	& $T_\mathrm{eff}$ & $\upsilon_\infty$	& Ref.\\
		&		&	&	[$\kms$] & [pc] & [$\spy$] &[K]& [$\kms$]\\
\hline
\multicolumn{2}{l}{\quad\it M-type stars}\\
\object{R Hor}	&	02:53:52.77	&	$-$49:53:22.7	&	37	&	310	&	$	5.9\e{-7}	$& 2200	& 4 &1\\
\object{IK Tau}	&	03:53:28.87 	&	+11:24:21.7 	&	34	&	265	&	$	5.0\e{-6}	$	& 2100 & 17.5 & 2\\
\object{GX Mon}	&	06:52:46.91	&	+08:25:19.0	&	$-$9	&	550	&	$	8.4\e{-6}	$	& 2600 & 19 &1\\
\object{W Hya}	&	13:49:02.00 	&	$-$28:22:03.5	&	40.5	&	78	&	$	1.0\e{-7}	$	& 2500 & 7.5 & 3\\
\object{RR Aql}	&	19:57:36.06	&	$-$01:53:11.3	&	28	&	530	&	$	2.3\e{-6}	$	& 2000 & 9 &1\\
\object{V1943 Sgr}	&	20:06:55.24	&	$-$27:13:29.8	&	$-$15	&	200	&	$	9.9\e{-8}	$	& 2200& 6.5 &1\\
\object{V1300 Aql}	&	20:10:27.87	&	$-$06:16:13.6	&	$-$18	&	620	&	$	1.0\e{-5}	$	& 2000 &14  &1\\
\object{V1111 Oph}   &   18:37:19.26 &  +10:25:42.2 &  $-$32 & 750 & $1.2\e{-5}$ & 2000 & 17 &1\\
\object{WX Psc}   &    01:06:25.98  &+12:35:53.1   &   9.5  & 700 & $4.0\e{-5}$ & 1800 & 19 & 5\\
\object{IRC -30398} & 18:59:13.85 & $-$29:50:20.4 & $-$6.5 & 550 & $8.0\e{-6}$ & 1800 &16 & 5\\
\multicolumn{2}{l}{\quad\it S-type stars}\\
\object{T Cet}	&	00:21:46.27	&	$-$20:03:28.9	&	22	&	240	&	$	6.0\e{-8}	$	& 2400 & 7 & 5\\
\object{TT Cen}	&	13:19:35.02	&	$-$60:46:46.3	&	4	&	880	&	$	4.0\e{-6}	$	& 1900 & 20 & 5\\
\object{W Aql}	&	19:15:23.35	&	$-$07:02:50.4	&	$-$23	&	395	&	$	4.0\e{-6}	$	& 2300 & 16.5 & 4\\
\object{RZ Sgr}	&	20:15:28.41	&	$-$44:24:37.5	&	$-$31	&	730	&	$	3.0\e{-6}	$	& 2400& 9 & 6\\
\multicolumn{2}{l}{\quad\it Carbon stars}\\
\object{R Lep}	&	04:59:36.35	&	$-$14:48:22.5	&	11	&	432	&	$	8.7\e{-7}	$	& 2200 & 18 &1\\
\object{V1259 Ori}	&	06:03:59.84	&	$+$07:25:54.4	&	42	&	1600	&	$	8.8\e{-6}	$	& 2200 & 16  &1\\
\object{AI Vol}	&	07:45:02.80	&	$-$71:19:43.2	&	$-$39	&	710	&	$	4.9\e{-6}	$	& 2100 &12 &1\\
\object{X TrA}	&	15:14:19.18	&	$-$70:04:46.1	&	$-$2	&	360	&	$	1.9\e{-7}	$	& 2200 & 6.5 &1\\
\object{II Lup}	&	15:23:04.91	&	$-$51:25:59.0	&	$-$15.5	&	500	&	$	1.7\e{-5}	$	& 2400 & 21.5  &1\\
\object{V821 Her}	&	18:41:54.39	&	+17:41:08.5	&	$-$0.5	&	600	&	$	3.0\e{-6}	$	& 2200 & 13.5 &1\\
\object{RV Aqr}	&	21:05:51.68	&	$-$00:12:40.3	&	1	&	670	&	$	2.3\e{-6}	$	& 2200 & 15 &1\\
\hline
\end{tabular}
\end{center}
\tablefoot{References give details of mass-loss rate, $\dot{M}$, stellar effective temperature, $T_\mathrm{eff}$, and distances. (1) \cite{Danilovich2015a}; (2) \cite{Maercker2016}; (3) \cite{Khouri2014} and \cite{Danilovich2016}; (4) \cite{Danilovich2014}; (5) \cite{Ramstedt2014}; (6) \cite{Schoier2013}.}
\end{table*}

\begin{table}[tp]
\caption{Observed transitions}\label{trans}
\begin{center}
\begin{tabular}{ccccc}
\hline\hline
Line & Frequency & Telescope &$\theta$	\\
 & [GHz] & & [$\arcsec$]\\
\hline
o-\h2S ($1_{1,0}\to1_{0,1}$)	&	168.763 & APEX & 37\\
p-\h2S ($2_{2,0}\to2_{1,1}$)	&	216.710 & APEX & 29\\
o-\h2S ($3_{3,0}\to3_{2,1}$)	&	300.506 & APEX & 21\\
 & & SMA & 1\\
o-\h2S ($3_{1,2}\to2_{2,1}$)& 1196.012\phantom{1} & HIFI & 19.5\\
o-\h2\up{34}S ($1_{1,0}\to1_{0,1}$) & 167.911 & APEX & 37\\
\hline
\end{tabular}
\end{center}
\tablefoot{$\theta$ is the HPBW}
\end{table}

\begin{table*}[tp]
\caption{Main beam integrated intensities for detections from APEX survey.}\label{APEXobs}
\begin{center}
\begin{tabular}{ccccc}
\hline\hline
 & \multicolumn{3}{c}{\h2S} & \h2\up{34}S\\
Star	&	$1_{1,0}\to1_{0,1}$	&	$2_{2,0}\to2_{1,1}$	&	$3_{3,0}\to3_{2,1}$	&	$1_{1,0}\to1_{0,1}$	\\
& [K $\kms$]& [K $\kms$]& [K $\kms$]& [K $\kms$]\\
\hline
{IK Tau}	&	1.85	&	0.26:	&	x	&	x	\\
{GX Mon}	&	1.07	&	x	&	x	&	x	\\
{V1300 Aql}	&	1.50	&	x	&	0.26:	&	0.34:	\\
V1111 Oph & 0.79 & ... & ... & x\\
WX Psc& 2.46 & ... & ... & 0.55\\
\hline
\end{tabular}
\end{center}
\tablefoot{(:) indicates a tentative detection; x indicates a non-detection (see Table \ref{nondet} for RMS); (...) indicates lines which were not observed.}
\end{table*}

\begin{table*}[tp]
\caption{RMS noise for all observed \h2S lines from APEX survey.}\label{nondet}
\begin{center}
\begin{tabular}{ccccc}
\hline\hline
 & \multicolumn{3}{c}{\h2S} & \h2\up{34}S\\
Star	&	$1_{1,0}\to1_{0,1}$	&	$2_{2,0}\to2_{1,1}$	&	$3_{3,0}\to3_{2,1}$	&	$1_{1,0}\to1_{0,1}$	\\
\hline
\multicolumn{5}{c}{\it M-type stars}\\
R Hor	&	11	&	...	&	13	&	11	\\
IK Tau & ~~18$^D$ & ~~17$^T$ & 20 & 18 \\
GX Mon	&  ~~20$^D$ &	12 &	18 &  20\\
W Hya	&	16	&	...	&	11	&	16	\\
RR Aql	&	19	&	15	&	13	&	19	\\
V1943 Sgr	& 13	&	...	&	11	&	13	\\
V1300 Aql & ~~13$^D$ & 20 & ~~15$^T$ & ~~13$^T$\\
V1111 Oph & ~~13$^D$& ... & ... & 13\\
WX Psc & ~~8$^D$& ... & ... & ~~8$^D$\\
IRC -30398 & 13 & ... & ... & 13\\
\hline
\multicolumn{5}{c}{\it S-type stars}\\
T Cet	&	...	&	13	&	...	&	...	\\
TT Cen	&	20	&	...	&	10	&	20	\\
W Aql	&	13	&	14	&	14	&	13	\\
RZ Sgr	&	20	&	...	&	13	&	20	\\
\hline
\multicolumn{5}{c}{\it Carbon stars}\\
R Lep	&	...	&	16	&	...	&	...	\\
V1259 Ori	&	17	&	14	&	15	&	17	\\
AI Vol	&	20	&	17	&	...	&	20	\\
X TrA	&	...	&	13	&	...	&	...	\\
II Lup	&	~~10*	&	11	&	10	&	...	\\
V821 Her	&	19	&	18	&	...	&	19	\\
RV Aqr	&	...	&	15	&	...	&	...	\\
\hline
\end{tabular}
\end{center}
\tablefoot{RMS values given in mK at a velocity resolution of $1~\kms$. (...) indicates lines which were not observed, ($^D$) and ($^T$) indicate detected or tentatively detected lines, respectively (see Table \ref{APEXobs} for integrated intensities). (*) indicates that the APEX observation is from De Beck et al (in prep).}
\end{table*}

Of the three \h2S lines observed as part of this project, the ortho-\h2S ($1_{1,0}\to1_{0,1}$) line was the brightest and most likely to be detected of the three \h2S lines observed. Hence, we have determined that a non-detection of the ($1_{1,0}\to1_{0,1}$) line places a firmer upper limit on the possible \h2S abundance in each source than a non-detection of either the para-\h2S ($2_{2,0}\to2_{1,1}$) line or the ortho-\h2S ($3_{3,0}\to3_{2,1}$), both of which were, at best, only tentatively detected in the sources that had clear ($1_{1,0}\to1_{0,1}$) detections.

\begin{figure}[t]
\includegraphics[width=0.5\textwidth]{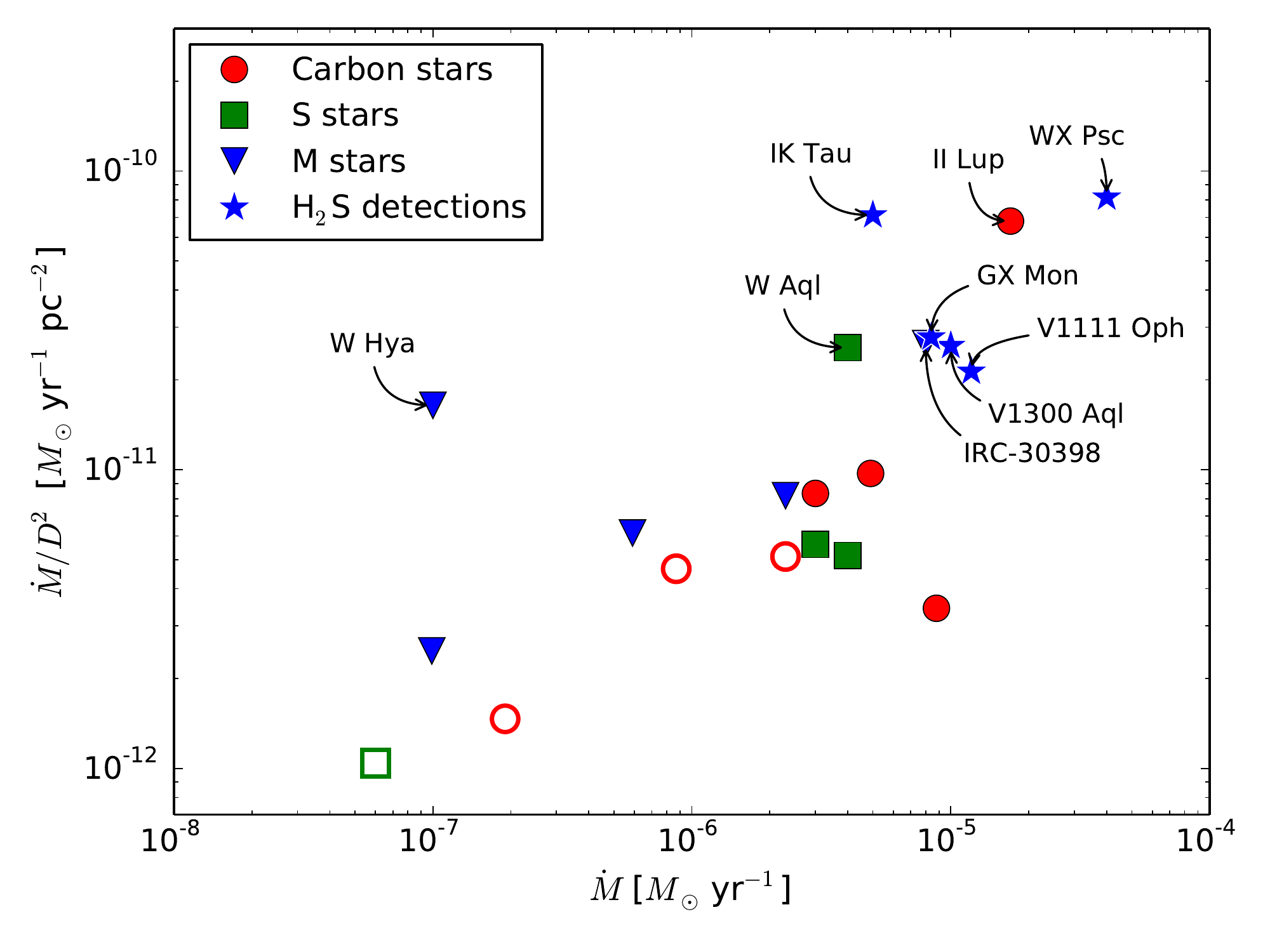}
\caption{A {predictor} of integrated line brightness, $\dot{M}/D^2$, plotted against mass-loss rate for our sample. Marker {types} denote which observed \h2S lines were undetected. {\it {Filled} markers} indicate that ($1_{1,0}\to1_{0,1}$) was undetected as well as the other two lines if they were observed. {\it {Unfilled} markers} indicate that ($2_{2,0}\to2_{1,1}$) was undetected and that the brighter ($1_{1,0}\to1_{0,1}$) and ($3_{3,0}\to3_{2,1}$) were not observed.}
\label{detnondet}
\end{figure}

{A first-order approximation for expected relative integrated line intensities between different stars is given by the mass-loss rate divided by the square of the distance, since integrated line intensity is generally expected to increase with mass-loss rate (neglecting specific abundances and optical depth effects) and decrease with the inverse square of the distance. In Fig. \ref{detnondet} this quantity, $\dot{M}/D^2$, is plotted against the mass-loss rate, $\dot{M}$,}
for all observed sources. As can be clearly seen, the sources for which \h2S has been detected all have high mass-loss rates and {are expected to exhibit} bright emission lines. However, \h2S was seen only in the M-type stars and not, for example, in the bright, high mass-loss rate carbon star II~Lup nor in the slightly less bright and slightly lower mass-loss rate S-type star W~Aql. 
The only non-M-type AGB star for which \h2S has been detected to date is CW~Leo which has a similar mass-loss rate to II Lup and, due to its proximity, a brightness of $1.4\e{-9}~\msol$~yr$^{-1}$~pc\up{-2}, putting it outside of our axes in Fig. \ref{detnondet}. We have also distinguished the non-detections with different markers depending on which lines were observed for each source and hence how strong the upper limit constraints we can place on \h2S abundances are. The {filled} markers indicate a non-detection of the strong ($1_{1,0}\to1_{0,1}$) line, while the {unfilled} markers indicate the non-detection of the ($2_{2,0}\to2_{1,1}$) and ($3_{3,0}\to3_{2,1}$) line in the absence of observations of the much brighter ($1_{1,0}\to1_{0,1}$) lines.

\begin{figure*}[t]
\centering
\includegraphics[width=1.0\textwidth]{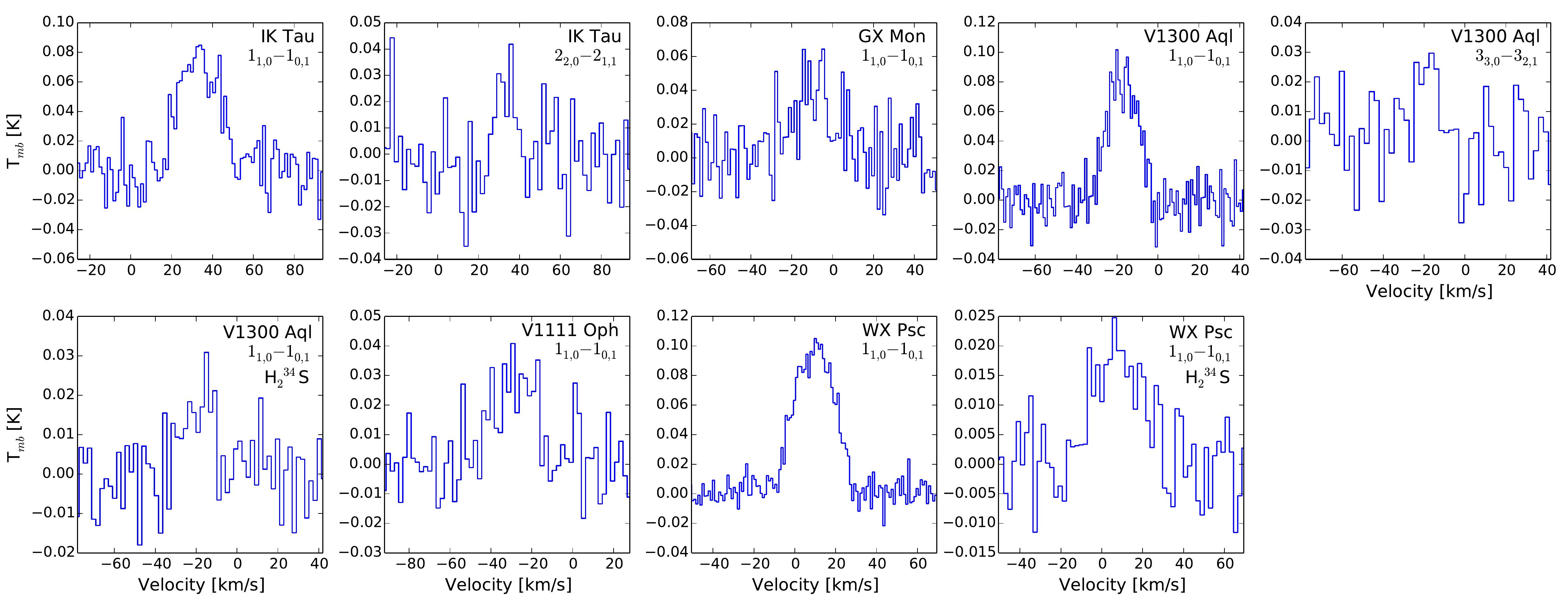}
\caption{All detections and tentative detections of \h2S and \h2\up{34}S lines from our APEX survey. All lines are marked with their transition (see Table \ref{trans} for frequencies). Observations of \h2\up{34}S lines are indicated below the transition numbers and all other lines are of \h2\up{32}S.}
\label{detlines}
\end{figure*}

In addition to the three targeted \h2\up{32}S lines, we concurrently observed the \h2\up{34}S ($1_{1,0}\to1_{0,1}$) line at 168.911 GHz and detected it in WX~Psc and, tentatively, in V1300~Aql. All detected \h2S lines are plotted in Fig. \ref{detlines} at a velocity resolution of 2~$\kms$, for the noisiest detections, or 1~$\kms$ for the clearer detections.

\subsection{Supplementary observations} 

The $(3_{3,0}\to3_{2,1})$ o-\h2S transition at 300.506 GHz was observed towards IK~Tau as part of an unbiased survey in the 279--355 GHz range using the Submilimetre Array (SMA) in the extended configuration. The full details of this survey are given in \cite{De-Beck2013}. {The size of the synthetic beam at this frequency was  $0.95\arcsec\times 0.93\arcsec$ and the \h2S emission was not resolved.} For our study, the emission line of interest was extracted for a synthetic circular beam 1\arcsec{} in diameter and is used to constrain our models. 

IK~Tau and WX~Psc (aka IRC+10011) were observed with \textsl{Herschel}/HIFI as part of the HIFISTARS guaranteed time key project \citep{de-Graauw2010,Bujarrabal2011,Roelfsema2012}. The o-\h2S ($3_{1,2}\to2_{2,1}$) line at 1196.012 GHz was observed as part of this project but not detected in either of the two sources in our sample \citep{Justtanont2012}. However, we use the non-detections as upper limits to help constrain our models. We re-reduced the data using the Herschel Interactive Processing Environment \citep[HIPE\footnote{\url{http://www.cosmos.esa.int/web/herschel/data-processing-overview}} version 14.2.1,][]{Ott2010} and the updated main beam efficiencies of Mueller et al. (2014)\footnote{\url{http://herschel.esac.esa.int/twiki/pub/Public/HifiCalibrationWeb/HifiBeamReleaseNote\_Sep2014.pdf}} which were released subsequent to the \cite{Justtanont2012} publication.

%
%
%
%
%
%


\section{Modelling}\label{modelling}

\subsection{Established parameters}\label{modparam}

Most of the stars in our survey were chosen from the sample presented in \cite{Danilovich2015a}. In that study, a large sample of AGB stars were observed across several CO emission lines and radiative transfer analyses were performed to determine circumstellar parameters based on the CO emission. Our circumstellar models are based on the results found in that paper. For WX~Psc, which was not included in \cite{Danilovich2015a}, we use the circumstellar model results from \cite{Ramstedt2014} and Danilovich et al. (\textit{in prep}), and for IK~Tau we use the results from \cite{Maercker2016}, all of which were obtained using the same {Monte Carlo} method used by \cite{Danilovich2015a}{, ensuring homogeneity. This modelling method was first described in detail by \cite{Schoier2001} and its reliability has been detailed by \cite{Ramstedt2008}, who note uncertainties up to a factor of $\sim3$. \cite{Danilovich2015a} also compare their results with past studies and conclude that on average they find mass-loss rates 40\% lower than past studies, most likely due to the modelling of of an acceleration region and their inclusion of relatively high-$J$ CO lines.} Some of the key circumstellar parameters for our sample --- LSR velocity, $\upsilon_{\mathrm{LSR}}$, distance, and mass-loss rate, $\dot{M}$ --- are listed in Table \ref{fullsample}.

\begin{figure*}[t]
\sidecaption
\includegraphics[width=0.7\textwidth]{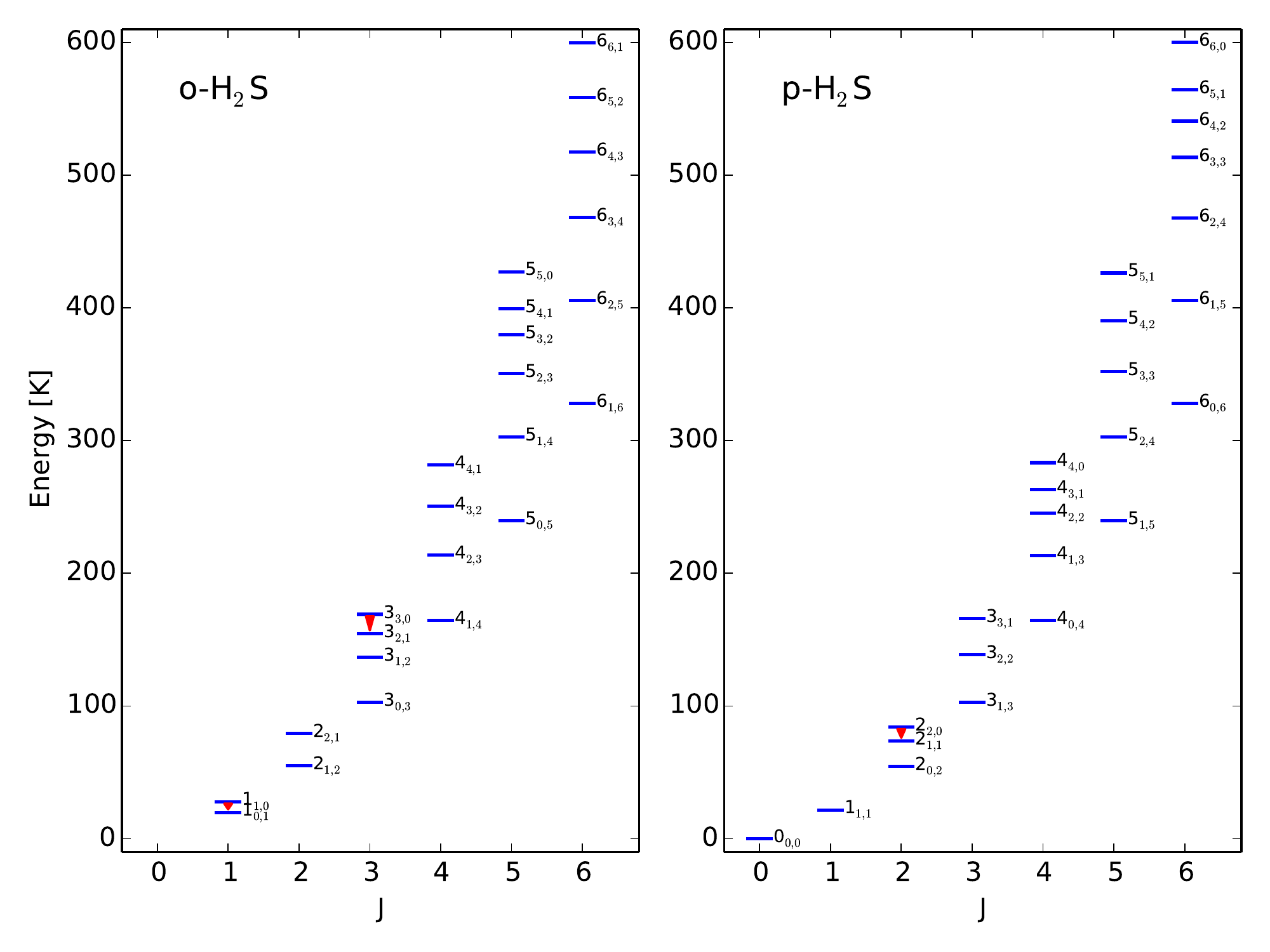}
\caption{An energy level diagram for \h2S with ortho energy levels shown on the left and para energy levels shown on the right. The quantum numbers are listed to the right of each level in the format $J_{K_a,K_c}$. {The red wedges indicate the transitions observed by APEX as part of this study.}}
\label{eld}
\end{figure*}

For the radiative transfer analysis we treated ortho and para \h2S separately since there are no gas-phase transitions linking the two spin isomers. We tested two molecular data files: one including only the ground vibrational state and collisional rates taken from \cite{Dubernet2009}, which was obtained from the LAMDA\footnote{\url{http://home.strw.leidenuniv.nl/~moldata/}} database \citep{Schoier2005}, and which we will henceforth refer to as the LAMDA description; and a more comprehensive description which includes vibrationally excited energy levels and collisional rates from \cite{Faure2007}, which we will refer to as the JHB description. The LAMDA description comprises 45 energy levels each for ortho- and para-\h2S and 139 and 140 radiative transitions for ortho- and para-\h2S, respectively, all taken from the JPL\footnote{\url{https://spec.jpl.nasa.gov}} spectroscopic database {\citep{Pickett1998}}. The collisional rates, which cover 990 collisional transitions for each of ortho- and para-\h2S at temperatures ranging from 5 to 1500 K, are scaled from the \h2O rates from \cite{Dubernet2009} and assume an \h2 ortho-to-para ratio (OPR) of 3. The JHB molecular data file contains 243 and 247 energy levels for ortho- and para-\h2S, respectively, with levels with energies up to approximately 6000 K. Levels included are rotational energy levels in the ground vibrational state up to $J_{K_a, K_c} = 12_{4,9}$ (ortho) and $J_{K_a, K_c} = 12_{3,9}$ (para), and levels in the $(\nu_1,\nu_2, \nu_3)$ = (0,1,0), (0,2,0), (0,3,0), (1,0,0), (0,0,1), (1,1,0), (0,1,1) vibrationally excited states up to and including all rotational levels with $J = 6$. The molecular data file also includes 1084 and 1039 radiative transitions for ortho- and para-\h2S, respectively, with all spectroscopic data taken from the HITRAN database \citep{Rothman2009}. The collisional rates cover temperatures from 20 to 2000 K and are scaled from the \h2O rates of \cite{Faure2007} and assume an \h2 OPR of 3. They cover energy levels from the ground state up to approximately 1000 K. 
In Fig. \ref{eld} we plot all ground state rotational energy levels up to $J=6$ for both ortho- and para-\h2S.


\subsection{Radiative transfer modelling procedure}

Based on the circumstellar parameters and molecular data files discussed above, we performed detailed radiative transfer modelling of the \h2S emission in our sources using an accelerated lambda iteration method code (ALI). ALI has been previously described in detail by \cite{Maercker2008} and \cite{Schoier2011} and has been implemented to model molecular emission from a wide variety of molecules in those studies and in  \cite{Danilovich2014,Danilovich2016}. Our models assume a smoothly accelerating and spherically symmetric CSE created by mass lost at a constant rate by the central AGB star. We assume \h2S abundance distributions based on chemical model results and on radial Gaussian profiles.

In the cases where more than one \h2S line was observed, we assume a Gaussian fractional abundance distribution of \h2S, {centred on the star and} described by 
\begin{equation}
f(r)= f_0 \exp\left(-\left(\frac{r}{R_e}\right)^2\right)
\end{equation}
where $f_0$ is the peak central abundance and $R_e$ is the $e$-folding radius at which the the abundance has dropped to $f_0/e$. As we have no \textit{a priori} constraints on the $e$-folding radius, we leave both $f_0$ and $R_e$ as free parameters in our modelling, to be adjusted to best fit the available data. Since we have some sources with only one \h2S line detected, we cannot properly constrain a Gaussian profile for those sources (see further discussion in Sect. \ref{abdisc}). In those cases we only use abundance distributions derived from chemical modelling (described further in Sect. \ref{chemmod}), scaled to fit our observations.

In general we have attempted to run our models using both the LAMDA and JHB molecular data files. We also extracted just the ground state information from the JHB description, to create the GS JHB description, to allow us to directly compare the impact of the vibrationally excited states on the excitation analysis of \h2S. Such a direct comparison cannot be done comparing the full JHB file with the LAMDA file as there may be additional variation introduced by the different collisional rates used in the two files.

Due to the uncertainty in the choice of molecular data file, we did not model our observations of para-\h2S, which was only tentatively detected towards IK~Tau, or \h2\up{34}S, which was detected towards WX Psc and tentatively detected towards V1300~Aql. The uncertainty in our o-\h2\up{32}S models (discussed in further detail below) is such that any comparisons would not be meaningful at present. A direct comparison of \h2\up{34}S with \h2\up{32}S based on line strengths is also not possible since the \h2\up{32}S ($1_{1,0}\to1_{0,1}$) line is optically thick for both V1300 Aql and WX Psc. Clearer and more numerous observations of p-\h2S and \h2\up{34}S would facilitate a more comprehensive and significant comparison. 


\subsection{Chemical modelling}\label{chemmod}

To put an additional constraint on the abundance distribution of \h2S for each of our stars, we derived abundance distributions from chemical modelling.
The forward chemistry model used is based on the UMIST Database for Astrochemistry (UDfA) CSE model. 
The chemical reaction network used is the most recent release of UDfA, \textsc{Rate12} \citep{McElroy2013}. The network includes only gas-phase reactions with a total of 6173 reactions involving 467 species. 
Both the UDfA CSE model and \textsc{Rate12} are publicly available\footnote{\url{http://udfa.ajmarkwick.net/index.php?mode=downloads}}.

The one-dimensional model assumes a uniformly expanding and spherically symmetric outflow, with both a constant mass-loss rate and expansion velocity, which are taken from the values listed in Table \ref{fullsample}. Self-shielding of CO is taken into account. A more detailed description can be found in \citet{Millar2000}, \citet{Cordiner2009}, and \citet{McElroy2013}.
We changed the {kinetic} temperature structure of the outflow to a power-law,
\begin{equation}
T(r) = T_* \left( \frac{R_*}{r} \right)^{-\epsilon},
\end{equation}
with $T_*=2000$~K, the stellar temperature, $R_*=5\e{13}$~cm, the stellar radius, and $\epsilon=0.65$, the exponent characterising the power-law.
In order to prevent unrealistically low temperatures in the outer CSE, the temperature has a set lower limit of 10 K \citep{Cordiner2009}.

The initial abundance of all species is assumed to be zero, except for that of the parent species. The parent species are injected into the outflow at the start of the model at  $1 \times 10^{14}$ cm.
We have calculated models using two lists of parent species: the O-rich list of \citet{Agundez2010} and the IK~Tau-specific list of \citet{Li2016}, {both based on observational constraints where available and supplemented with thermal equilibrium calculations}. These two model results are henceforth referred to as the A and L abundance distributions after the sources of the parent species lists. 

Inputting these abundance distributions as calculated into our radiative transfer models resulted in severe under-predictions for all our observed \h2S lines. To tune these models so that they agreed with the observations, we scaled the derived abundance distributions until we found a model that best fit the data. This approach is justified since we found that increasing the initial (parent) abundance of \h2S in the forward chemistry models had the effect of increasing the abundance distribution by a uniform scale factor.


%

\subsection{Modelling results}

For each source we have found best fit models for each of the L and A radial abundance distributions coupled with each of the available molecular data files: LAMDA, JHB and GS JHB. This gives six models for each source. In addition, for IK~Tau and V1300~Aql, the sources with multiple detections, we also found three best fit models for a Gaussian abundance distribution paired with each of the molecular data files. 

\begin{figure*}[t]
\centering
\includegraphics[width=0.49\textwidth]{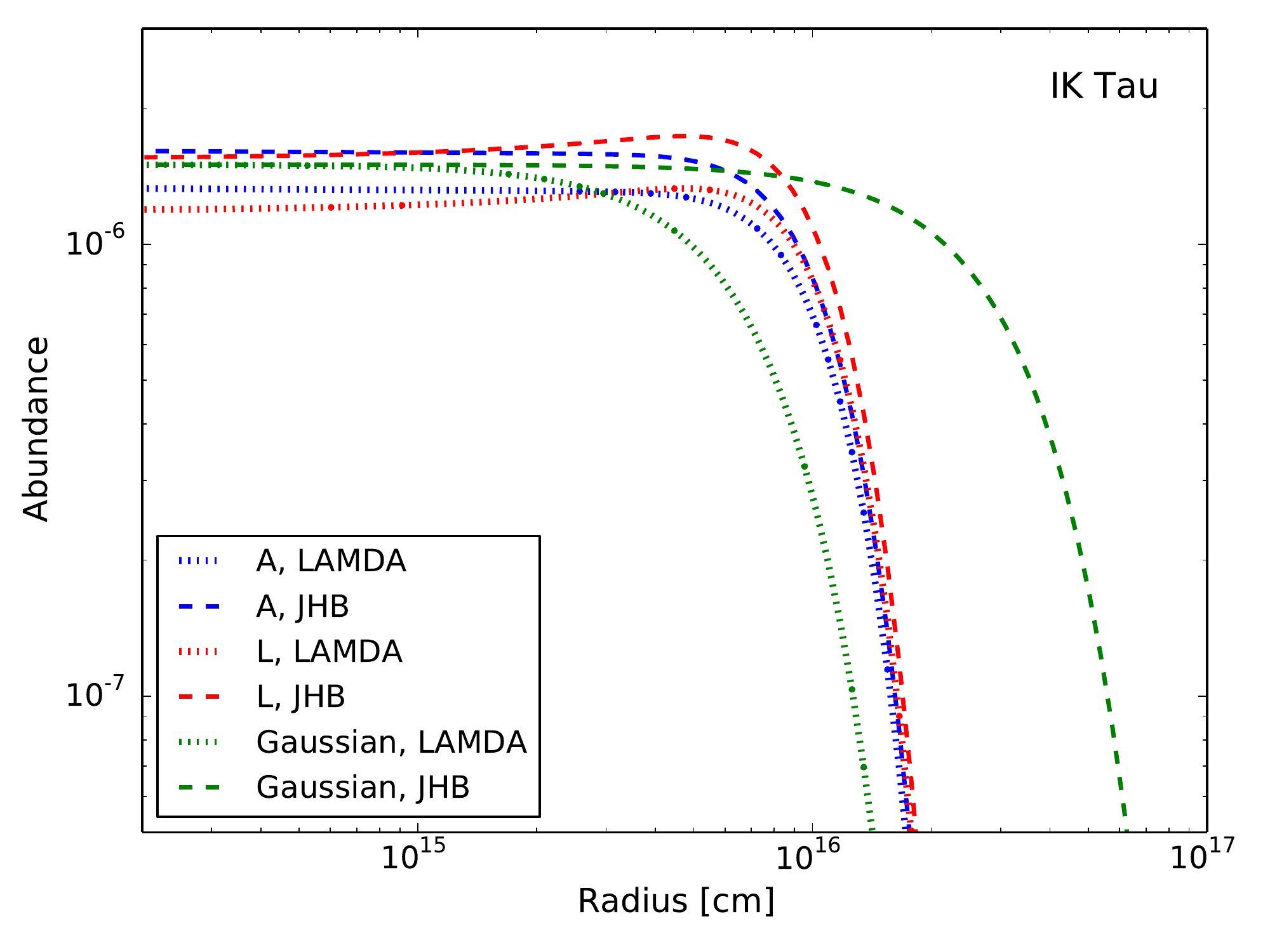}
\includegraphics[width=0.49\textwidth]{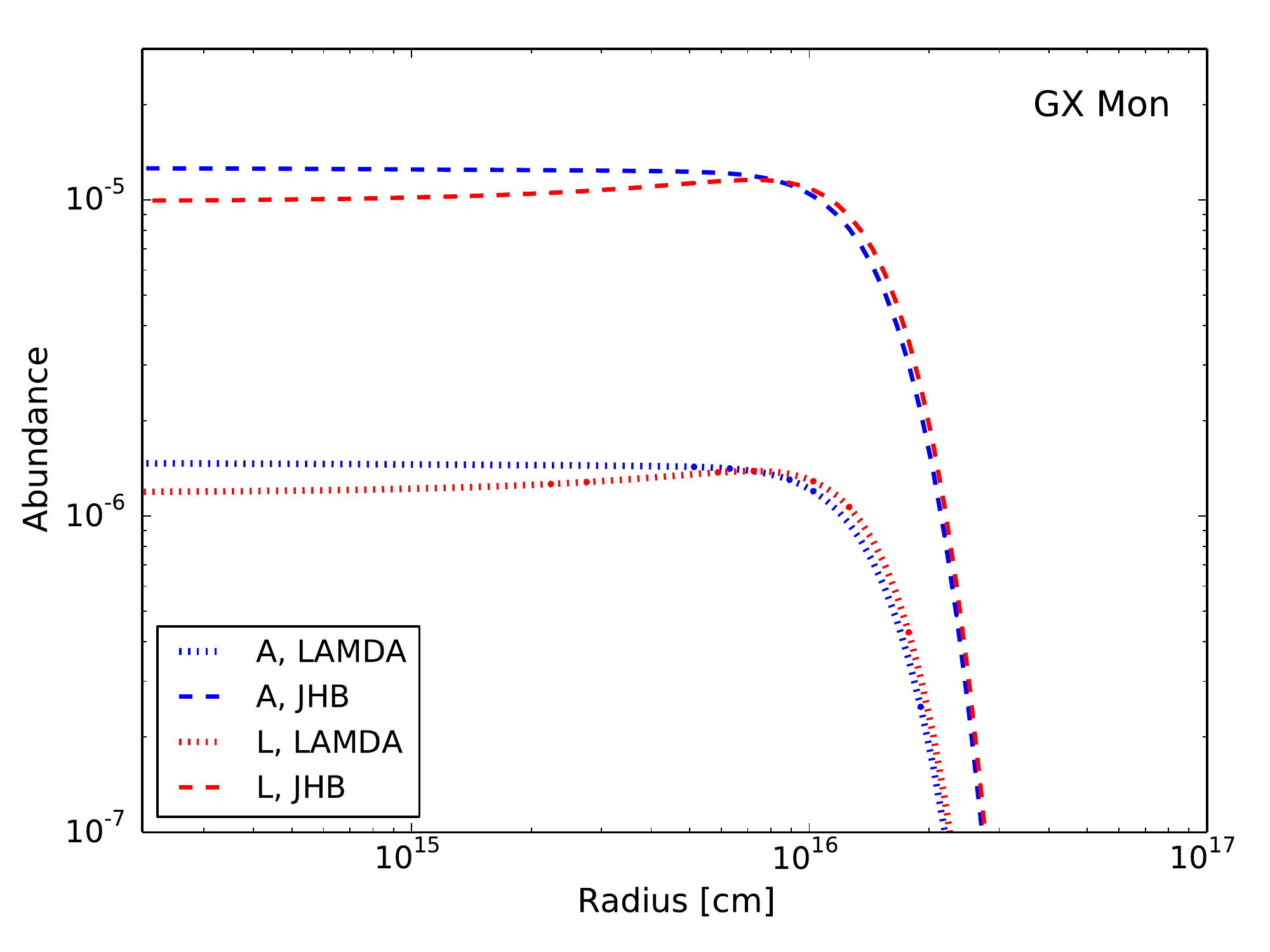}
\includegraphics[width=0.49\textwidth]{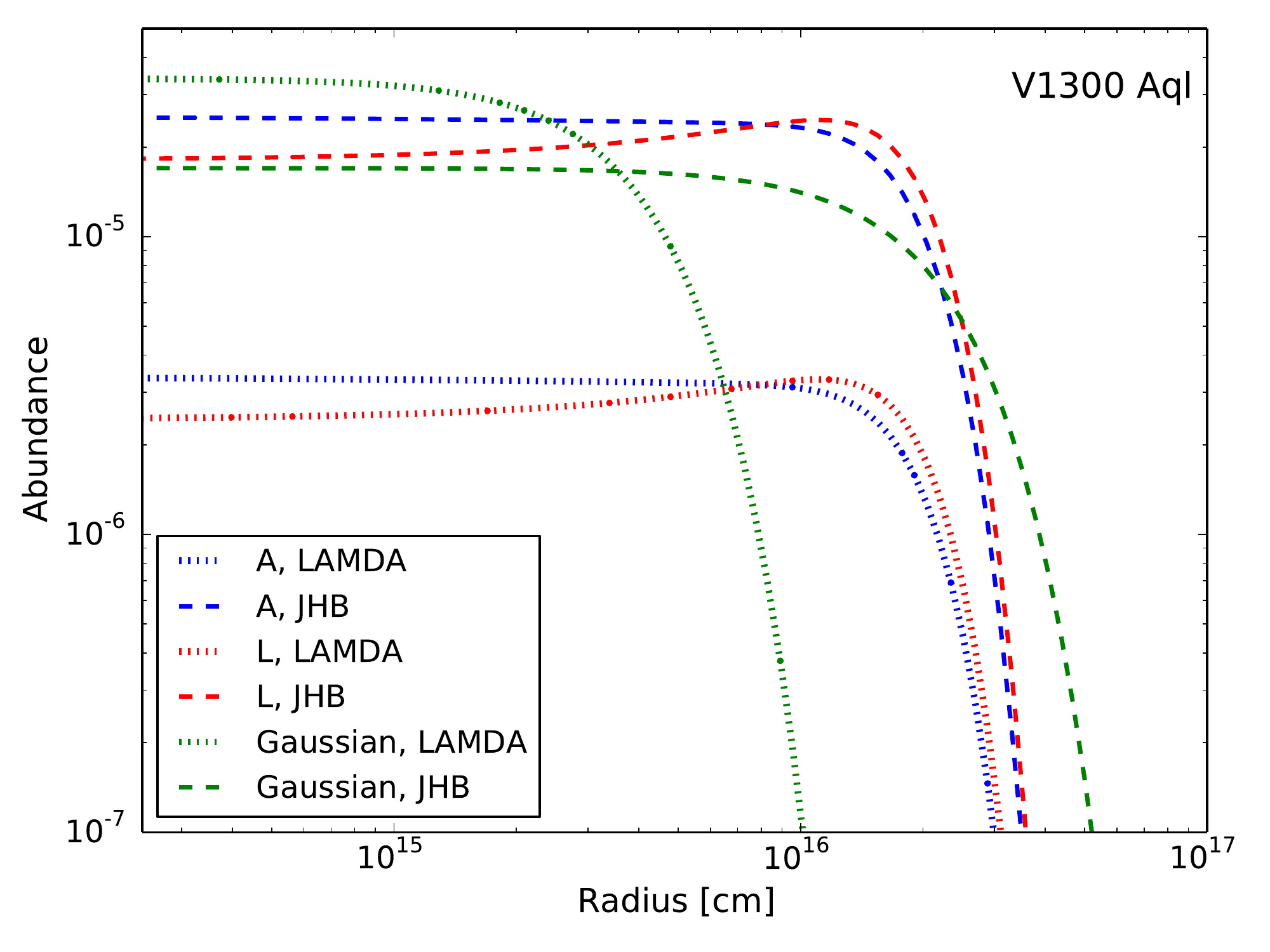}
\includegraphics[width=0.49\textwidth]{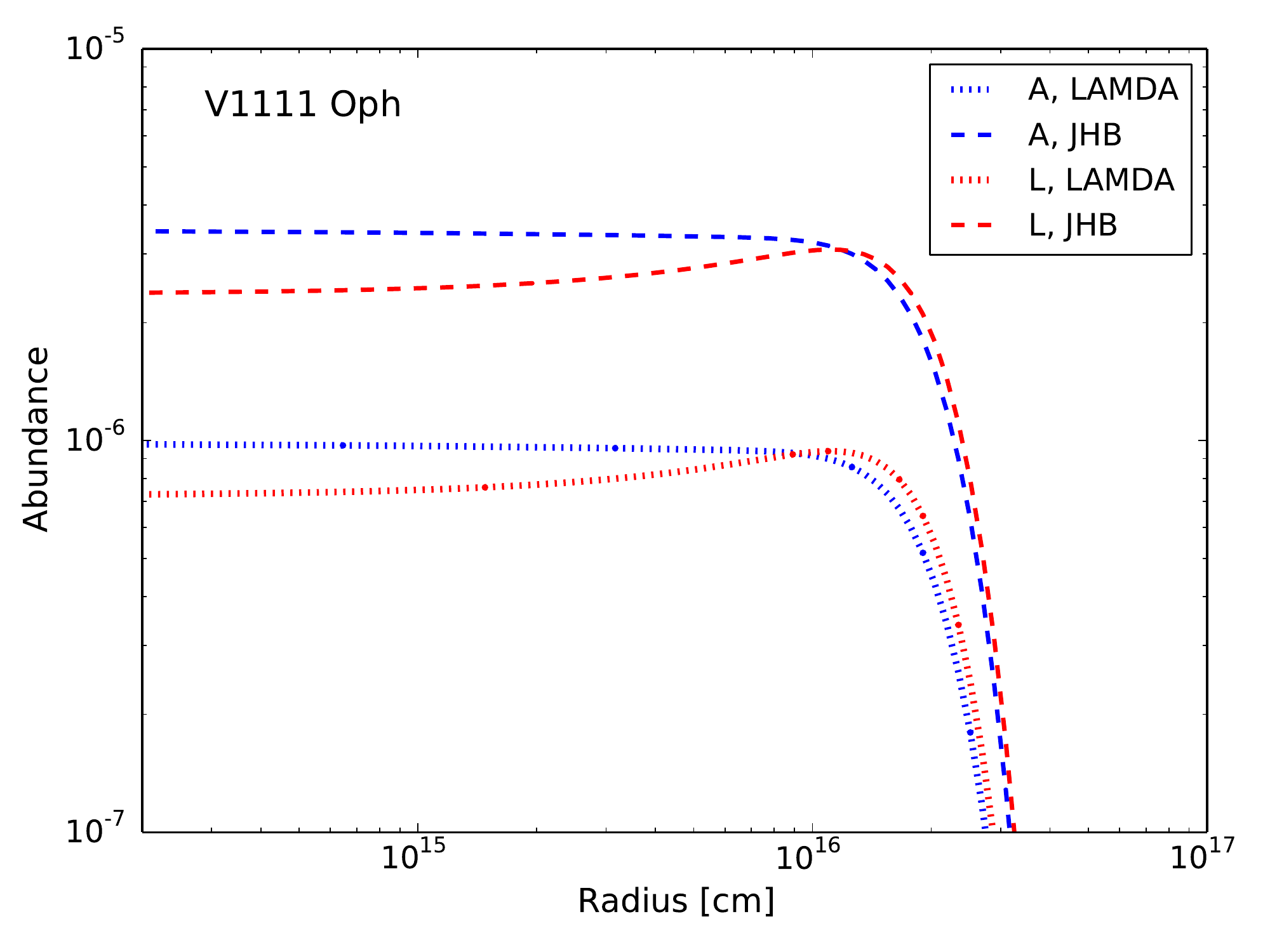}
\includegraphics[width=0.49\textwidth]{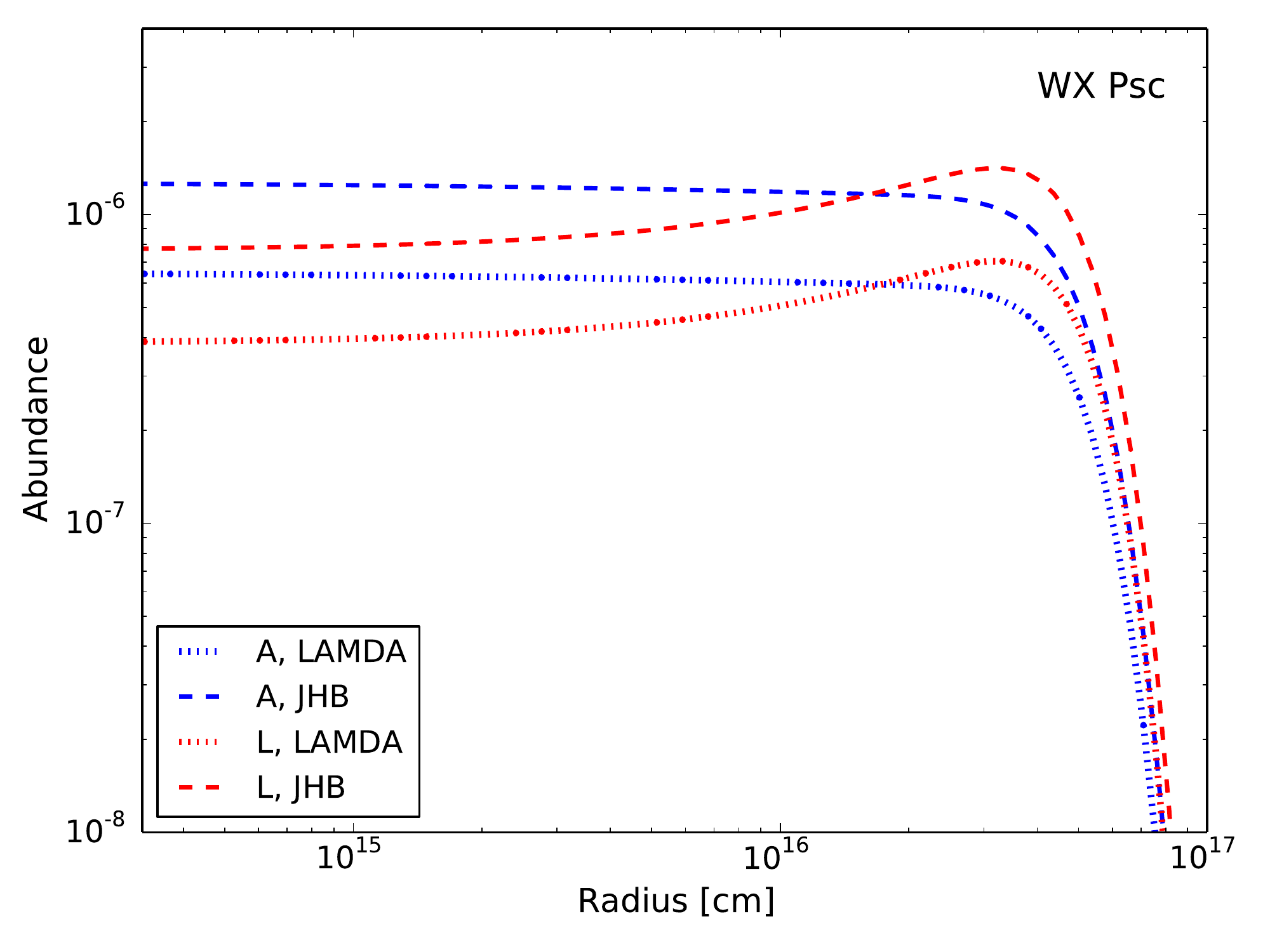}
\caption{Plots of abundance distributions, o-\h2S/\h2, that best fit our data with different molecular data files. Dotted lines represent models using the LAMDA molecular data file, dashed lines represent models using the JHB molecular data file. Blue curves are abundance distributions based on chemical models arising from parent molecule list A, red curves similarly arise from parent molecule list L, and green curves are Gaussian abundance distributions.}
\label{abundanceresults}
\end{figure*}

\begin{figure}[t]
\centering
\includegraphics[width=0.5\textwidth]{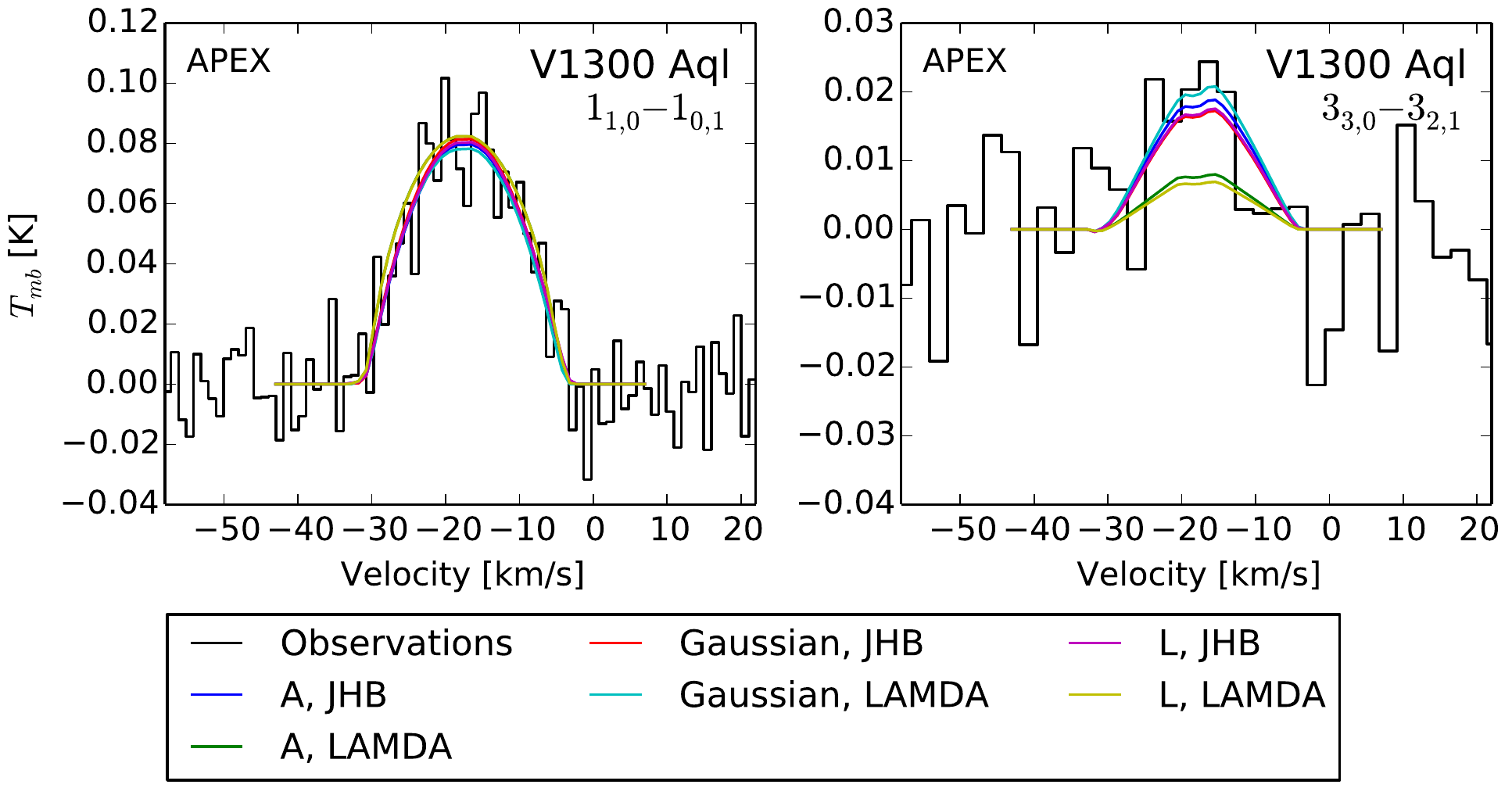}
\includegraphics[width=0.5\textwidth]{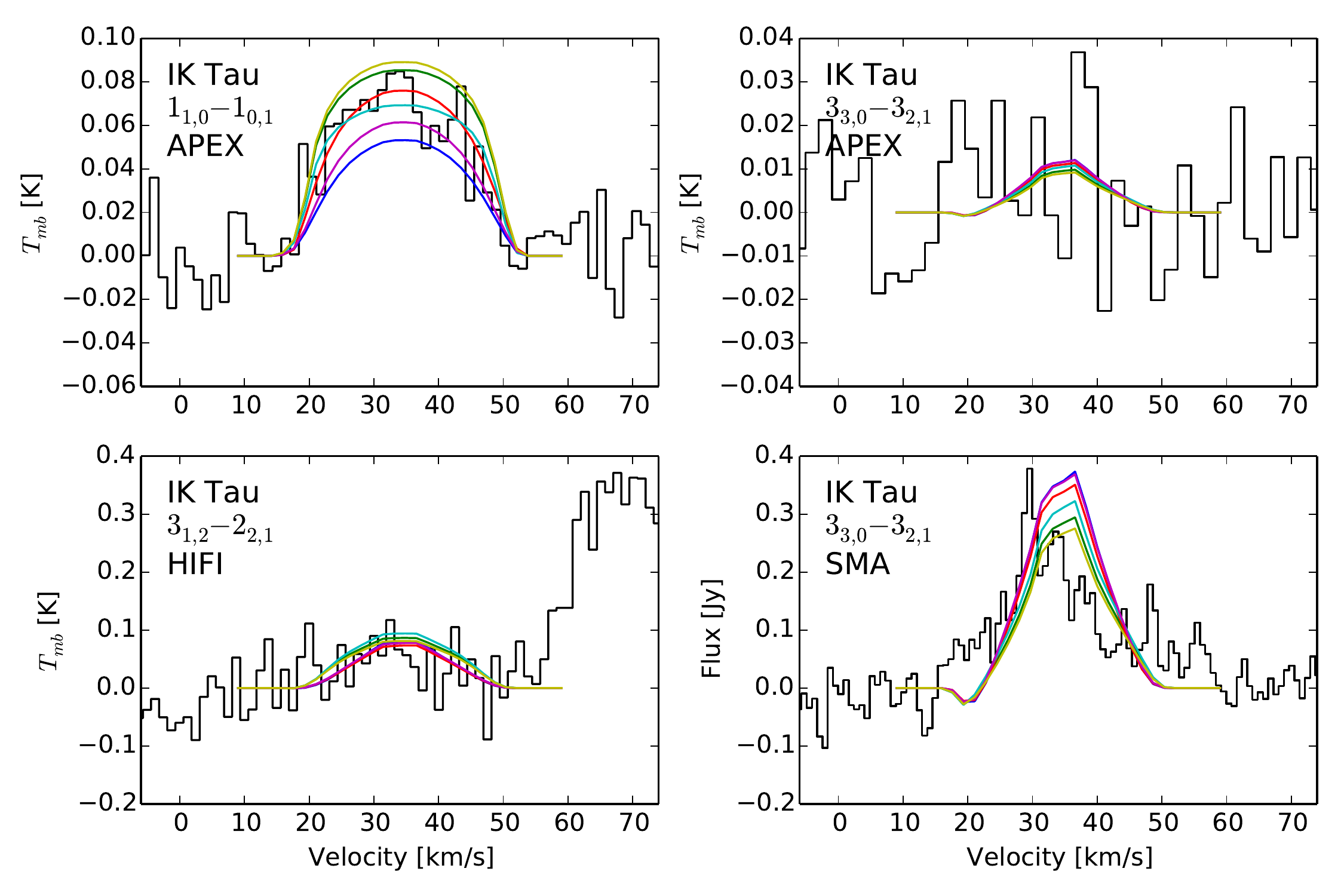}
\caption{Observations and model lines for o-\h2S towards V1300~Aql (top) and IK~Tau (bottom). Different coloured curves indicate the results of best-fit models derived using different methods described in the text.}
\label{resultplot}
\end{figure}

The derived abundances and radii for our sources are listed in Table \ref{results} and the abundance distributions that best match our data (either scaled from chemical models or derived Gaussian abundance profiles) are plotted in Fig. \ref{abundanceresults}. The model results for V1300~Aql and IK~Tau are plotted in Fig. \ref{resultplot}. These two stars were chosen as examples because they are the only two for which we have two detected lines (and additionally two undetected lines for IK~Tau), and hence were the only two stars for which we could calculate models with Gaussian abundance distributions. In Fig. \ref{resultplot} we show six different variations on the choice of abundance distribution and molecular excitation file. For V1300~Aql, for which there is a strong detection of the ($1_{1,0}\to1_{0,1}$) line and a tentative detection of the ($3_{3,0}\to3_{2,1}$) line, the models are most strongly constrained by the stronger line, since the weaker line has a larger uncertainty. This is particularly evident when considering that all the models fit the ($1_{1,0}\to1_{0,1}$) line equally well, whereas the two models with chemically modelled abundances paired with the LAMDA molecular data file predict much weaker ($3_{3,0}\to3_{2,1}$) emission than the other four models. From this we can conclude that, for V1300~Aql, the models with fixed abundance distributions --- which are similar in radial size to the Gaussian distribution we find for the JHB molecular data file --- do not give realistic results when paired with the LAMDA molecular data file. This effect is more pronounced for V1300~Aql than for IK~Tau, for which we find generally lower o-\h2S abundances.

Overall, as can be seen in Table \ref{results} and Figs. \ref{abundanceresults} and \ref{resultplot}, the different molecular data files give varying results --- most notably different peak abundances --- when other factors such as the abundance distribution were held constant. Conversely, changing the abundance distribution induced less significant differences in peak abundances when the molecular data files were held constant. These effects are discussed in more detail in Sections \ref{exdisc} and \ref{abdisc}. {We have not included formal uncertainties for our models since these are much smaller than the difficult-to-quantify errors introduced by the choice of collisional rates.}

%
%
%

\begin{table}[tp]
\caption{Modelling results for o-\h2S.}\label{results}
\begin{center}
\begin{tabular}{r|ccc}
\hline\hline
&LAMDA & JHB & GS JHB\\
\hline
\multicolumn{1}{l|}{IK Tau} \\
L scale: & 36 & 48 & 51\\
 $f_\mathrm{peak}$ & $1.1\e{-6}$ & $1.7\e{-6}$ & $1.9\e{-6}$\\ 
A scale: & 19 & 23 & 25\\
 $f_\mathrm{peak}$ & $1.1\e{-6}$ & $1.6\e{-6}$ & $1.8\e{-6}$\\ 
Gaussian: $f_0$ & $1.5\e{-6}$ & $1.5\e{-6}$ & $1.6\e{-6}$\\
$R_e$ [cm] & $7.7\e{15}$ & $3.4\e{16}$ & $3.4\e{16}$\\
\hline
\multicolumn{1}{l|}{GX Mon} \\
L scale: & 36 & 300 & 410\\
 $f_\mathrm{peak}$ & $1.4\e{-6}$ & $1.2\e{-5}$ & $1.6\e{-5}$\\ 
A scale: & 21 & 180 & 250\\
 $f_\mathrm{peak}$ & $1.5\e{-6}$ & $1.3\e{-5}$ & $1.8\e{-5}$\\ 
\hline
\multicolumn{1}{l|}{V1300 Aql} \\
L scale: & 74 & 550 &800\\
 $f_\mathrm{peak}$ & $3.3\e{-6}$ & $2.5\e{-5}$ & $3.6\e{-5}$\\ 
A scale: & 48 & 360 & 380\\
$f_\mathrm{peak}$ & $3.4\e{-6}$ & $2.5\e{-5}$ & $2.7\e{-5}$\\ 
Gaussian: $f_0$ & $3.4\e{-5}$ & $1.7\e{-5}$ &$2.7\e{-5}$\\
\phantom{Gaussian:} $R_e$ [cm] & $4.2\e{15}$ & $2.3\e{16}$ &$2.3\e{16}$\\
\hline
\multicolumn{1}{l|}{V1111 Oph} \\
L scale: & 22 & 72 & 77\\
 $f_\mathrm{peak}$ & $9.4\e{-7}$ & $3.1\e{-6}$ & $3.3\e{-6}$\\ 
A scale: & 14 & 49 & 52\\
 $f_\mathrm{peak}$ & $9.8\e{-7}$ & $3.4\e{-6}$ & $3.6\e{-6}$\\ 
\hline
\multicolumn{1}{l|}{WX Psc} \\
L scale: & 12 & 24 & 26\\
 $f_\mathrm{peak}$ & $7.1\e{-7}$ & $1.4\e{-6}$ & $1.5\e{-6}$\\ 
A scale: & 9.2 & 18 & 19\\
 $f_\mathrm{peak}$ & $6.4\e{-7}$ & $1.3\e{-6}$ & $1.3\e{-6}$\\ 
\hline
\end{tabular}
\end{center}
\tablefoot{$f_0$ is the peak abundance relative to \h2 and $R_e$ is the $e$-folding radius. The scales for Li and Agundez refer to the multiplicative scale factor required to fit the abundance distributions produced from chemical models using the L and A parent molecules, respectively, and $f_\mathrm{peak}$ is the peak abundance relative to \h2 after scaling those abundance distributions. See text for further details.}
\end{table}


\section{Discussion}

\subsection{The choice of molecular excitation data}\label{exdisc}


In examining the results of using different molecular data files we determined that the most significant difference was not in the inclusion or not of the vibrationally excited states --- although this did play a small role as can be seen when comparing the results of the JHB description and the GS JHB description --- but in the choice of collisional rates. Note also that both molecular data files only include collisional rates for transitions in the ground vibrational state.

The LAMDA file includes collisional rates based on those determined by \cite{Dubernet2006,Dubernet2009} and \cite{Daniel2010,Daniel2011} for \h2O collisions with p-\h2 and o-\h2, scaled to the mass of \h2S. 
The JHB file includes collisional rates based on the \h2O rates calculated by \cite{Faure2007}, which are based on quasiclassical trajectory calculations and are most suited to high-temperature situations. Indeed, when discussing \h2O, \cite{Dubernet2009} recommends the use of the \cite{Faure2007} collisional coefficients for high temperatures above 400 K, cautioning that the weakest transitions included by \cite{Faure2007} are based on scaled \h2O-He collisions, which are the least accurate. Below 400~K, \cite{Dubernet2009} and \cite{Daniel2011} recommend the use of their own collisional rates. However, the temperatures within AGB CSEs range from cool ($\sim 10$ K) to warm ($\sim 2000$ K), spanning both sides of the suggested 400 K cutoff. The review of \cite{van-der-Tak2011} suggests that for \h2O the collisional rates of \cite{Dubernet2006} are the best to use for low temperatures ($\leq 20$ K, such as found in some molecular clouds), while at warm temperatures ($\sim 300$ K) the best collisional rates are those from \cite{Faure2007}, and at high temperatures ($\geq 300$ K) the collisional rates from \cite{Faure2008}, which include vibrational excitation, are preferred. This does not, however, avoid the problem of AGB CSEs spanning all three temperature regimes.

One of the key problems here is that the precise correspondence of the \h2O rates to the \h2S rates is not presently known. Merely scaling to account for the different masses of \h2O and \h2S does not take into account differences in molecular cross-section and dipole moments. Furthermore, \cite{Daniel2011} note that the \h2O-\h2 rates obtained by scaling rates from \h2O-He \citep[such as those by][]{Green1993} are the least reliable. There is no similar data on the reliability of scaling \h2O-\h2 rates to \h2S-\h2 rates, which may be significant, given that the dipole moments of the two molecules differ by a factor of $\sim2$ (1.8546 D for \h2O \citep{Lide2003} and 0.974 D for \h2S \citep{Viswanathan1984}).

Another issue when deriving \h2S collisional rates from \h2O collisional rates is that the distribution of energy levels for \h2S is not perfectly analogous to that of \h2O. 
We note that for the JHB molecular data file these subtleties are taken into consideration, whereas for the LAMDA description they are not, so that for some transitions the de-excitation rate is instead listed as the excitation rate (because of a reversed energy order with respect to the \h2O levels), slightly altering the model.

\begin{figure*}[t]
\sidecaption
\includegraphics[width=0.7\textwidth]{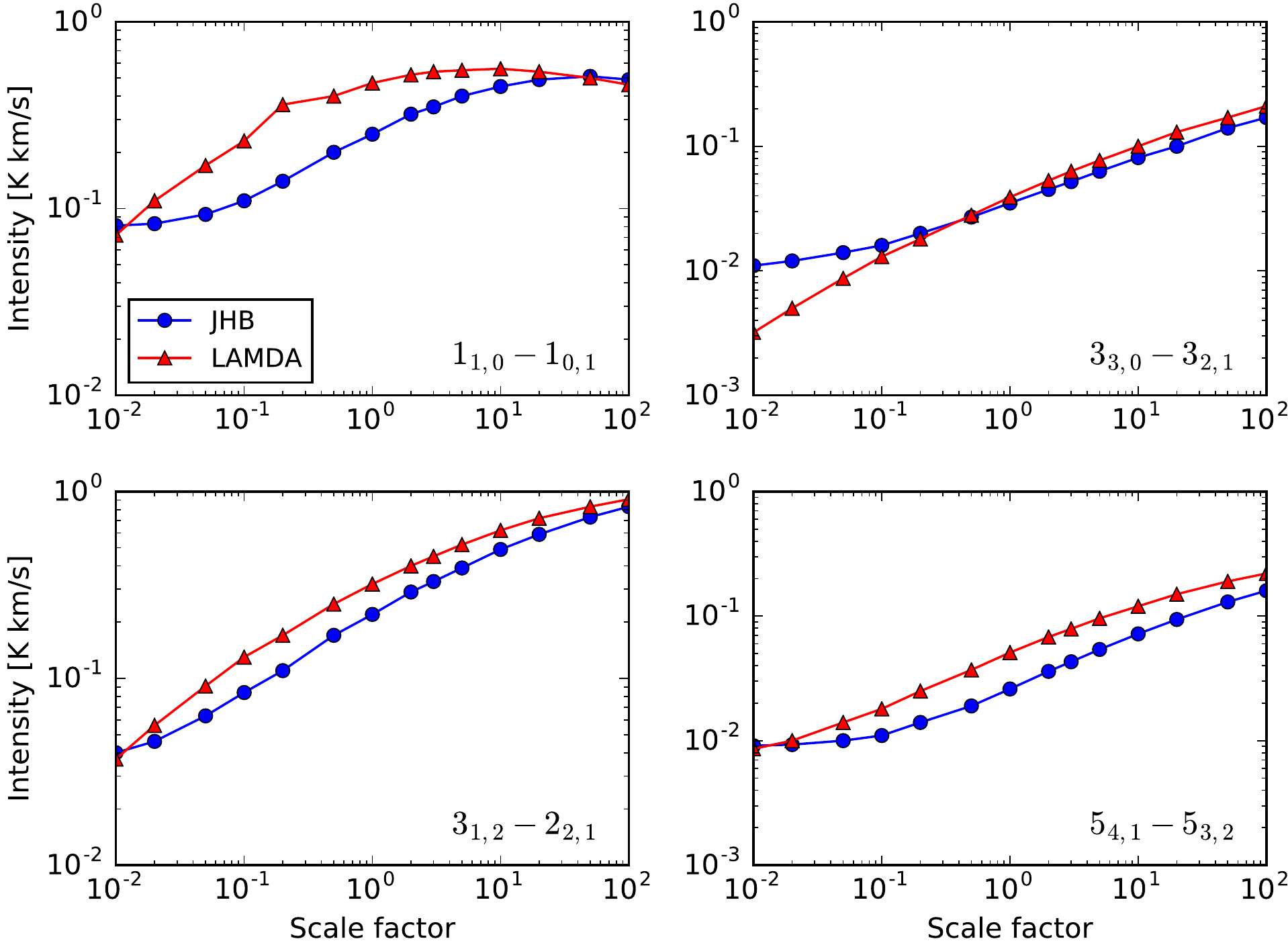}
\caption{The variation in integrated line intensities for four \h2S transition lines as a result of uniform scaling of the collisional rates given in the LAMDA and JHB molecular data files (see text for further details).}
\label{colrates}
\end{figure*}

To further quantify the dependence of \h2S on the choice of collisional rates, we scaled the collisional rates given in both the JHB and the LAMDA files by several factors up to two orders of magnitudes in both increasing and decreasing directions. All rates were scaled uniformly and the resulting comparison was made for the modelled integrated intensity for four o-\h2S transition lines, the three key lines from our various observations: ($1_{1,0}\to1_{0,1}$), ($3_{3,0}\to3_{2,1}$), and ($3_{1,2}\to2_{2,1}$), and a higher-$J$ line at 407.677 GHz, ($5_{4,1}\to5_{3,2}$).  For these test models we used the basic parameters for V1111~Oph as listed in Table \ref{fullsample} and Sect. \ref{modelling}. We assumed a Gaussian \h2S distribution with an inner \h2S abundance of $1\e{-6}$ and an $e$-folding radius of $1\e{16}$~cm. The results, showing the variation in integrated line intensity with collisional rate scale factor, are plotted in Fig. \ref{colrates}. Varying the collisional rates as we have done has a significant impact on the resulting integrated line intensities. This strongly suggests that \h2S is primarily collisionally excited rather than radiatively excited, and hence that the choice of collisional rates is important to precisely constrain its abundance and distribution in a CSE. The choice of collisional rates appears to have a larger impact than the smaller range of energy levels and radiative transitions included in the LAMDA file compared with the JHB file. Similar tests of collisional rates for \so2 resulted in very minor changes in integrated line intensities, implying that \so2 is instead primarily radiatively excited \citep{Danilovich2016}. Unlike for \so2, the importance of the collisional rates for \h2S indicates that better rate determinations, measured or calculated directly for \h2S rather than scaled from rates for \h2O, are likely to significantly improve the accuracy of radiative transfer models of \h2S.

Furthermore, as can be seen in the top two plots of Fig. \ref{colrates}, there is a more significant difference in final integrated line intensity between the the two molecular data files for the ($1_{1,0}\to1_{0,1}$) line than for the ($3_{3,0}\to3_{2,1}$) line, when considering the unscaled sets of collisional rates. This suggests that better detections of a wide variety of lines --- those which do vary significantly with the choice of collisional rates as well as those that do not --- could help discriminate between choices of molecular data file. Since the biggest difference between the Gaussian abundance profiles for the best models with each of the two molecular data files was the $e$-folding radius, spatially resolved observations, such as those which can be performed with the Atacama Large Millimetre/submillimetre Array (ALMA), will not only allow us to put better constraints on the \h2S distribution but also on the choice of molecular data file. Neither of these, however, is a full substitute for updated collisional rates calculated specifically for \h2S.

\subsubsection{The effects of excluding excited vibrational states.}\label{vibex}

From the results tabulated in Table \ref{results}, it can be seen that the peak abundances required for the models to fit our observations did not change very significantly between the JHB and GS JHB molecular data files. However, there was a tendency for the ground state file to require larger abundances than the full file to reproduce the observed results. A similar effect was seen for NH$_3$ by \cite{Schoier2011} and \cite{Danilovich2014}, although the effect is less pronounced for \h2S. 
These results do not conclusively justify the exclusion of the vibrationally excited states, although the choice of collisional rates does play a more significant role in altering the model results.

\subsection{Choice of abundance distribution}\label{abdisc}

The different abundance distributions used in our best-fit models are plotted in Fig. \ref{abundanceresults}. There it can be seen that there are only minor differences between the abundance distributions resulting from the chemical models using the A and L parent molecules, for a given molecular data file.

\begin{figure*}[t]
\centering
\includegraphics[width=0.78\textwidth]{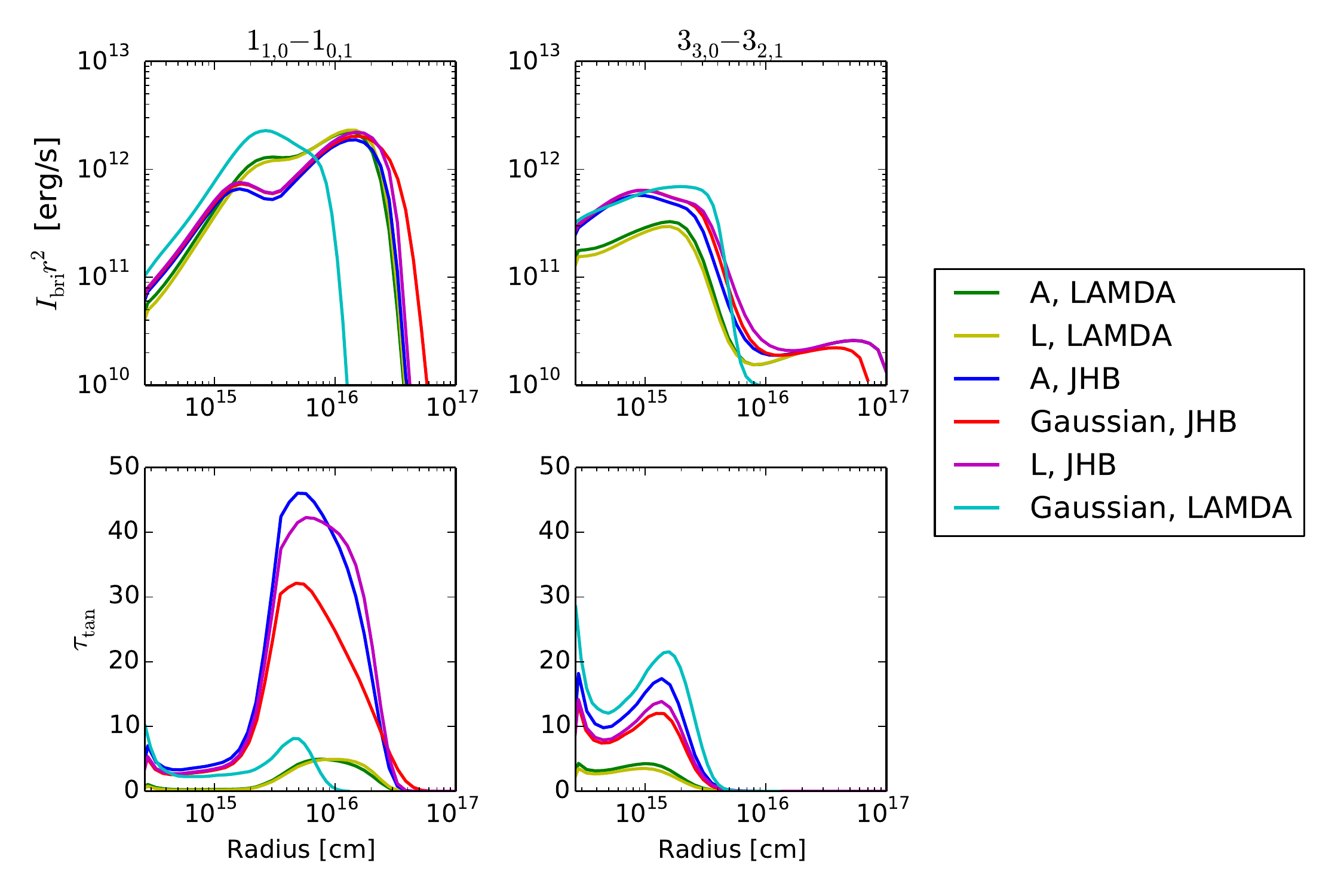}
\includegraphics[width=0.8\textwidth]{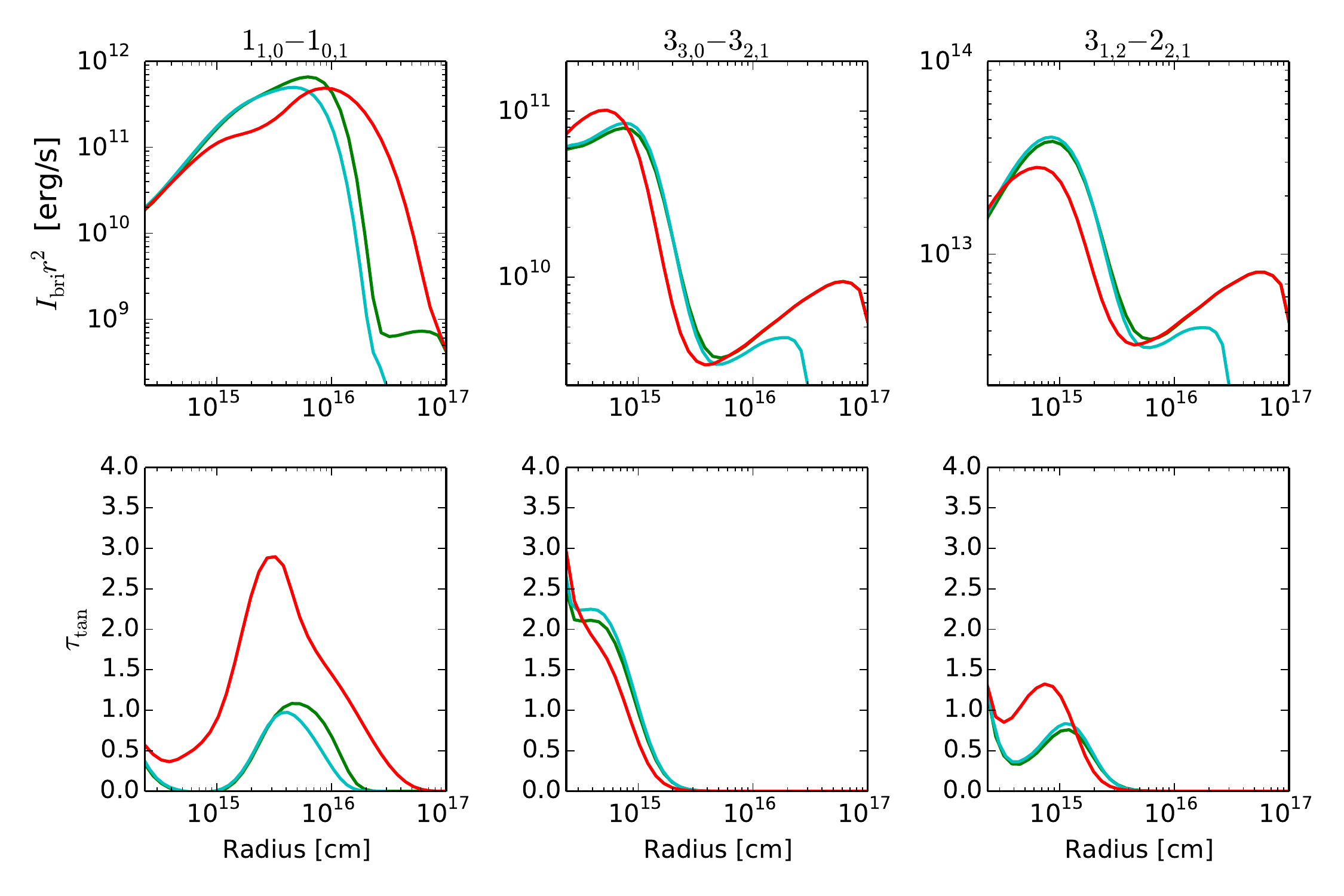}
\caption{Results from various model fits to o-\h2S for V1300~Aql (top) and IK~Tau (bottom). Transition numbers at the top of a column indicate the line for which the parameters are plotted. The top row of each star's plots shows the {tangential} brightness distribution of the emission line, {$I_\mathrm{bri}$, scaled with the radius squared, $r^2$, to emphasise the features more clearly, plotted} against radius. The corresponding bottom row shows the tangential optical depth of the emission line with radius.  The legend in the top right is given for V1300~Aql and also applies for the more sparsely selected lines for IK~Tau.}
\label{emittingregions}
\end{figure*}

In the case of IK~Tau and V1300~Aql we had access to two o-\h2S detections from various telescopes. The two observed lines, ($1_{1,0}\to1_{0,1}$) and ($3_{3,0}\to3_{2,1}$), have reasonably separated emitting regions, with the lower-$J$ line mostly emitting from the outer regions of the molecular envelope and the higher-$J$ line emitting more strongly from the inner regions of the molecular envelope, as seen in {the brightness distribution plots in} Fig. \ref{emittingregions}. Hence, when fitting a Gaussian abundance distribution, we left the $e$-folding radius as a free parameter along with the peak abundance and constrained both with our observations. This lead to large differences in radii between models using the LAMDA and JHB molecular data files, as seen in Fig. \ref{abundanceresults}.


For the sources with only one detection, it is not possible to put any constraints on a Gaussian abundance profile, hence our not including any such results in Table \ref{results}. The upper limit given by the non-detection for GX~Mon adds some constraints, but not enough to give sufficient certainty. This is why we only include model results based on the abundance distributions generated by chemical modelling for GX~Mon, WX~Psc, and V1111~Oph.

The strongest emission lines in the frequency range accessible from APEX (and most ground-based telescopes) are those that we observed for this project and a few lines that fall in the $\sim400$--500 GHz region. {These higher frequency lines} require long integration times to reach sufficiently good RMS noise levels to obtain clear detections {due to several key \h2S lines falling in regions of poor atmospheric transmission, even in good weather conditions}. Indeed, the p-\h2S ($1_{1,1}\to0_{0,0}$) line, {which is expected to be bright and} which would have provided an interesting point of comparison with the o-\h2S ($1_{1,0}\to1_{0,1}$) line, lies at 452.390 GHz, {in the wing of a strong water vapour absorption feature at 448 GHz\footnote{This can be clearly seen using the APEX atmospheric transmission calculator at \url{http://www.apex-telescope.org/sites/chajnantor/atmosphere/}}}. Observations in this region were unfortunately unsuitable for inclusion in the broad sulphur survey of which the present results are a part. However, now that we have some stars with clear \h2S detections, future targeted observations of the more accessible higher frequency \h2S lines would allow us to better constrain the \h2S emission in the observed AGB stars. The best and most reliable way to precisely constrain the abundance distribution of \h2S{, especially in the absence of space-based observations,} would be to use an interferometer capable of resolving the emitting region to observe a reliable \h2S line. ALMA will soon have the capability to observe the \h2S ($1_{1,1}\to0_{0,0}$) line at 168.763 GHz when the Band 5 receiver (covering 157--212 GHz) is available for general observing (expected for ALMA Cycle 5 from early-2018). {The sizes of the \h2S emitting regions predicted by our models are on the order of a few arcseconds, depending on the source, well within ALMA's resolving capability.}

\subsection{Comparison with other studies}

Previous observations of \h2S in AGB stars have mainly yielded detections towards OH/IR stars. \cite{Omont1993}, for example, surveyed 34 sources and detected \h2S towards all seven OH/IR stars, four out of nine M-type AGB stars (other than the OH/IR stars), one out of three carbon AGB stars, and neither of the two S-type AGB stars. Our detection pattern is in general agreement with their results (see Fig. \ref{detnondet}), aside from our exclusion of extreme OH/IR stars. Four of our five stars were also observed by \cite{Omont1993} who detected the same 168 GHz \h2S line in all four. Additionally, they found a weak detection of the 216 GHz \h2S line towards WX Psc, a line which was not observed as part of our survey.

\cite{Omont1993} also model the \h2S emission for WX~Psc and OH~26.5~+0.6 including infrared rotational excitation. They find a high abundance of \h2S $\sim 1\e{-5}$ for both stars, accounting for a significant fraction of the sulphur budget, and their abundance distribution is a step function out to $\sim 10^{16}$~cm, similar in size to many of our models. Their calculated abundance for WX Psc is an order of magnitude larger than our models, although they do note the significant uncertainty in their model due to noisy observations. They also find that models with smaller radial distributions of \h2S yield implausibly large abundances, which is in agreement with our modelling results.

The earlier study of \cite{Ukita1983} failed to detect \h2S in all but OH~231.8~+4.2, an extreme OH/IR star. Their source list includes lower mass-loss rate AGB stars (and some other types of objects) and there is little overlap with our source list, since they mainly observed northern sources and we mainly observed southern sources. 

\cite{Yamamura2000} identify the sulphur-bearing molecule HS in R~And, an S-type AGB star, using high resolution infrared spectra. They note that HS is located in the stellar atmosphere and moves inwards during stellar pulsations. They do not simultaneously detect \h2S in their spectra. Since R~And is a relatively low mass-loss rate S-type star \citep[$\dot{M} = 5\e{-7}\spy$,][]{Danilovich2015a}, we would not expect to detect \h2S based on the survey results in this present study. 

\subsection{Comparison with chemical models}

The L and A abundance distributions from chemical model results used in our radiative transfer modelling were based on the parent species listed in the \cite{Li2016} and \cite{Agundez2010} studies, and not on their direct results. For example, \cite{Li2016} base their models on the example of IK~Tau specifically and assume that the S-bearing parent species are \h2S, SO, \so2, CS and SiS. Their resultant radial distributions of these molecules start high at their inner radii (set to $10^{15}$~cm) and gradually decrease aside from a slight increase of \h2S which can be seen in our Fig. \ref{abundanceresults}. They find a much lower \h2S abundance than our radiative transfer results for IK~Tau (and indeed, all five of our sources), which are a few orders of magnitude higher. The \cite{Li2016} results are also in disagreement with the observational SO results found by \cite{Danilovich2016} for IK~Tau, which show a lower inner abundance of SO with a peak at $\sim 10^{16}$~cm.

The \cite{Agundez2010} study covers a similar region of the CSE --- beginning at $10^{14}$~cm, further inwards than \cite{Li2016} --- but primarily focuses on the effect of clumpiness in the CSE. Their S-bearing parent species for oxygen-rich CSEs are SiS, CS, and \h2S, with a much lower \h2S abundance than we find in any of our models, by at least two orders of magnitude. 

The models of \cite{Willacy1997} examine a similar region of the CSE as those of \cite{Li2016} but assume different parent molecules and corresponding initial abundances. Their parent molecules include SiS and \h2S, with a large initial abundance of \h2S that accounts for most of the sulphur in the CSE. They use TX~Cam as their example star, which is not in our sample but is similar to IK~Tau in terms of mass-loss rate. However, unlike our results for IK~Tau, their resultant \h2S distribution is in close agreement with our Gaussian and JHB model for V1300~Aql in terms of both abundance and distribution size.

The chemical models of \cite{Gobrecht2016} primarily look at the innermost regions of the CSE, within the dust condensation radius, and focus on shock-induced chemistry. Their model also uses IK~Tau as the exemplar source. Their initial abundances are based on thermal equilibrium calculations and they begin with a high abundance of \h2S, accounting for most of the sulphur. This drops off within 9 stellar radii to only $\sim 10^{-8}$, a few orders of magnitude lower than our models, which have inner radii close to the outer radius used by \cite{Gobrecht2016}.

From the differences in the above models, especially those which examine a similar region of the CSE as our radiative transfer models, it is clear that a different choice of parent molecules and their abundances can change the abundance distributions predicted by the chemical models. Since we found a range of o-\h2S abundances among our sources, from $4\e{-7}$ to $2.5\e{-5}$, it is also likely that different conditions (temperature, density, possibly age) in different stars lead to different abundances of various molecules such as \h2S. A goal of our ongoing work is to determine abundances for five key S-bearing molecules (\h2S, SO, \so2, CS, SiS) for a consistent sample of stars based on observations and radiative transfer modelling. These results can then be compared with chemical models \textit{en masse} and, ideally, adjusting the chemical models to agree with these results will yield more precise chemical models that better represent individual AGB stars.


\subsection{Trends, or lack there of, in H$_2$S abundance}

V1300~Aql was found to have a higher \h2S abundance than the other four stars in our sample, especially when considering the models calculated using the JHB molecular data file and/or a Gaussian abundance distribution. The \h2S abundance is about an order of magnitude higher than that found for V1111~Oph, the source with the most similar mass-loss rate, and more than an order of magnitude higher than that found for WX Psc, the source with the largest mass-loss rate, a factor of four higher than the mass-loss rate of V1300~Aql. The most similar abundance and mass-loss rate combination we found was towards GX~Mon, which has a slightly lower mass-loss rate than V1300~Aql, a slightly lower \h2S abundance and a similar separation between abundances calculated from the JHB and LAMDA molecular data files. However, the GX~Mon models are based primarily on only one \h2S detection, which is also noisier than the corresponding V1300~Aql transition, making those models slightly less certain.

There is no clear correlation between our modelled \h2S abundances --- when looking at a single, consistent modelling method --- and mass-loss rate or CSE density ($\dot{M}/\upsilon_\infty$). {The clearest correlation between mass-loss rate and \h2S abundance is based on the fact that \h2S was only detected for the highest mass-loss rate stars, as shown in Fig. \ref{detnondet}. As emphasised there,  \h2S was not detected towards W~Hya, a bright M-type star (with a similar brightness to V1111~Oph), which has a low mass-loss rate, tentatively suggesting that \h2S may only be present in higher mass-loss rate stars.
}
Outside of the sample of stars examined here, the most significant group of AGB stars with \h2S detections are the {high mass-loss rate} extreme OH/IR stars, such as those from the studies of \cite{Ukita1983}, \cite{Omont1993}, and \cite{Justtanont2015}. 
Of the stars in our sample, none have been confirmed to be extreme OH/IR stars. Indeed, extreme OH/IR stars were excluded from our sulphur survey sample due to the difficulty in finding a certain circumstellar model from CO emission lines that are often rife with interstellar contamination. 


\cite{Hashimoto1997} suggest that V1300~Aql has recently become an OH/IR star in the superwind phase \citep{Iben1983,Justtanont2013,de-Vries2014}, based on an analysis of the compact and thick circumstellar dust envelope. More recently, \cite{Cox2012} studied the extended dust emission of a large sample of AGB stars and report V1300~Aql to have no extended emission that can be resolved at 70 or 160~$\mic$ by \textsl{Herschel}/PACS photometry, which supports the premise of a small dust envelope. Furthermore, \cite{Ramstedt2014} find a low \up{12}CO/\up{13}CO ratio of 7, suggesting that V1300~Aql has undergone hot bottom burning \citep{Lattanzio2003}. Both the OH/IR phase and hot bottom burning occur at the end of a star's tenure on the AGB so the presence of one is suggestive of the other.

If V1300~Aql is indeed in the early stages of the superwind phase, this could explain why it has a higher \h2S abundance and hence that it should be grouped with the OH/IR stars in this respect. Although precise \h2S abundances are not presently know for extreme OH/IR stars, they are likely to be significant based on the intensity of various observations such as those by \cite{Omont1993}, who detected the 168 GHz and 216 GHz \h2S lines in all their surveyed OH/IR stars and whose analysis suggests abundances on the order of $\sim10^{-5}$.


There is not presently enough information to draw firm conclusions as to whether \h2S is more abundant in extreme OH/IR stars, let alone any explanations as to why that may be the case. Further investigation, as well as a more accurate molecular data file (see Sect. \ref{exdisc}), will allow us to better constrain the occurrence of \h2S in AGB stars.

\subsection{H$_2$S and ramifications for the sulphur budget}

Our results indicate that for some {oxygen-rich} AGB stars {with high mass-loss rates}, \h2S {may} account for a significant fraction of the sulphur budget. This is in contrast with SO and \so2, which were found by \cite{Danilovich2016} to be the most significant carriers of sulphur in low mass-loss rate oxygen-rich AGB stars. In carbon stars the most significant carriers of sulphur are CS and SiS \citep{Olofsson1993a,Schoier2007} with \h2S only playing a minor role, if any at all.

Since sulphur is not nucleosynthesised in AGB stars or their progenitors, we expect the overall sulphur abundance to match that of the solar neighbourhood, although with most of our sources at distances of 550--750 pc (except for IK~Tau at 265 pc) this may not necessarily hold. However, \cite{Rudolph2006} show that while there is a trend for higher sulphur abundances towards the galactic centre and lower abundances in the outer parts of the galaxy, it is not a steep gradient and the approximate S/\h2 $\sim 2-3\e{-5}$ that can be derived from their results is likely to hold for all of our sources. 
We find a slight overabundance of \h2S in two of our V1300~Aql models (the Gaussian + LAMDA model and the L + ground state JHB model), which give slightly more \h2S than the expected abundance of sulphur allows. We have already discussed the likely over-prediction of \h2S due to the exclusion of excited vibrational states and these are likely a symptom of that issue. Our remaining models for all sources do not yield higher \h2S abundances than the expected sulphur abundance. 

The only source for which the \h2S abundance is high enough to account for all of the sulphur is V1300~Aql, although for GX~Mon \h2S also accounts for a significant portion of the sulphur. Indeed, applying an ortho-to-para ratio of 3 to our results in Table \ref{results} pushes the abundances well into the expected  S/\h2 range (or above, in the cases noted above). {The ortho-to-para ratio of 3 comes from assuming statistical equilibrium (valid for warm formation temperatures) and, due to the low quality of our only para-\h2S detection, is not a ratio we can confirm observationally or through modelling.}
The concern here is the possibility of other S-bearing molecules also having significant abundances and consequently pushing the derived total sulphur abundance above realistic levels. Our subsequent papers in this series will investigate the abundances of the other key sulphur molecules observed towards V1300~Aql and GX~Mon in more detail, but a {preliminary} analysis of V1300~Aql gives peak SiS, SO, and \so2 abundances of $\sim 1\e{-6}$ each, and a peak abundance of CS of $\sim 1\e{-7}$. This brings the total sulphur abundances to within the considerable uncertainties of the expected value and suggests that other molecules such as HS and atomic S are likely to play, at most, only a minor role in the circumstellar envelope of V1300~Aql.

\section{Conclusions}

In this study, we present new \h2S observations acquired as part of a larger APEX survey of sulphur-bearing molecules in the circumstellar envelopes of AGB stars. Our clearest detections are of the 168.763 GHz o-\h2S ($1_{1,0}\to1_{0,1}$) line, which is the main focus of our subsequent analysis. We also tentatively detect the equivalent o-\h2\up{34}S line and one p-\h2S line. As well as the five sources with detected \h2S lines, we present a comprehensive series of non-detections with sensitive RMS limits for a further 16 stars of various chemical types and a range of mass-loss rates. 

We perform detailed radiative transfer models of o-\h2S for all five detected sources, using two abundance distributions based on the results of chemical models and, where a second o-\h2S line is detected, using a Gaussian distribution profile. For each model we use three different molecular data files, mostly differing in their choices of collisional de-excitation rates, which were scaled from different calculations of \h2O collisional rates. We find a spread of peak abundances of o-\h2S ranging from $\sim 4\e{-7}$ to $3\e{-5}$, depending on the source and the modelling method used. Overall, we determine that \h2S can account for a significant fraction of the overall sulphur budget, and we find it to be the dominant S-bearing molecule in V1300~Aql and GX~Mon in particular.

Because we have, at most, only one detection of p-\h2S or o-\h2\up{34}S for each source, we did not model these two species. Since the lines in question --- or the o-\h2\up{32}S lines being compared to --- are optically thick, it is not possible to make accurate comparisons of abundances or isotopologue ratios without full radiative transfer models. This will only be possible in the future if we can obtain more observations to allow for more constrained models of these species. Alternatively, more certain o-\h2S models with updated calculations of collisional rates will also allow for better constraints on \h2S in general and hence its isotopologues and spin isomers.

\begin{acknowledgements}

The authors would like to acknowledge John H Black for his compilation of the \h2S molecular data file used in our modelling.

LD and TD acknowledge support from the ERC consolidator grant 646758 AEROSOL and the FWO Research Project grant G024112N. MVdS and LD acknowledge support from the Research Council of the KU Leuven under grant number GOA/2013/012. TJM acknowledges support from the STFC, grant reference ST/P000312/1.

Based on observations made with APEX under programme IDs O-097.F-9318 and O-098.F-9305.

APEX is a collaboration between the Max-Planck-Institut f\"ur Radioastronomie, the European Southern Observatory, and the Onsala Space Observatory. 

HIFI has been designed and built by a consortium of institutes and university departments from across Europe, Canada and the United States under the leadership of SRON Netherlands Institute for Space Research, Groningen, The Netherlands and with major contributions from Germany, France and the US. Consortium members are: Canada: CSA, U.Waterloo; France: CESR, LAB, LERMA, IRAM; Germany: KOSMA, MPIfR, MPS; Ireland, NUI Maynooth; Italy: ASI, IFSI-INAF, Osservatorio Astrofisico di ArcetriNAF; Netherlands: SRON, TUD; Poland: CAMK, CBK; Spain: Observatorio Astron\'omico Nacional (IGN), Centro de Astrobiolog\'ia (CSIC-INTA). Sweden: Chalmers University of Technology -- MC2, RSS \& GARD; Onsala Space Observatory; Swedish National Space Board, Stockholm University -- Stockholm Observatory; Switzerland: ETH Zurich, FHNW; USA: Caltech, JPL, NHSC.

\end{acknowledgements}

%

\bibliographystyle{aa}
\bibliography{H2S}


\end{document}